\documentclass[preprint,showpacs,floats,letterpaper,floatfix,,groupedaddress,eqsecnum]{revtex4}
\usepackage{subfigure}

\usepackage{amssymb,amsmath}
\usepackage[dvips]{graphicx}

\usepackage{dcolumn,epsfig}

\begin{document}

\title{Stochastic Resonance in Periodic Potentials Revisited.}

\author{W.L. Reenbohn, S.S. Pohlong and Mangal C. Mahato}
\email{mangal@nehu.ac.in}
\affiliation{Department of Physics, North-Eastern Hill University, 
Shillong-793022, India}

\begin{abstract}
The phenomenon of Stochastic Resonance (SR) has been conclusively 
demonstrated in bistable potentials. However, SR in sinusoidal potentials have
only recently been shown numerically to occur in terms of hysteresis loop area.
We show that the occurrence of SR is not specific to sinusoidal potentials and
can occur in periodic bistable potential, $U(x)=\frac{2}{3}(\cos x +\cos 2x)$,
as well. We further show that SR can occur even in a washboard potential, where 
hysteresis loops normally do not close because of average drift of particles. 
Upon correcting for the drift term, the closed hysteresis loop area (input 
energy loss) shows the usual SR peaking behaviour as also the signal-to-noise 
ratio in a limited domain in the high drive-frequency range. The occurrence of 
SR is attributed to the existence of effectively two dynamical states in the
driven periodic sinusoidal and periodic bistable potentials. The same 
explanation holds also when the periodic potentials are tilted by a small 
constant slope.

\end{abstract}

\vspace{0.5cm}
\date{\today}

\maketitle

\section{Introduction}
The phenomenon of stochastic resonance (SR), initially put forth theoretically 
to explain the occurrence of ice ages, at a certain frequency, on 
earth\cite{Benzi}, have been observed experimentally in many 
physical\cite{Fauve, Roy} and biological systems\cite{Douglass, Moss}. SR has 
been investigated theoretically and experimentally with considerable interest 
over the past three decades\cite{Gamma, Well}. Its attraction lies in the 
seemingly counter-intuitive idea that by tuning noise level (externally or 
internally) the response of a nonlinear system to a weak external periodic 
signal can be enhanced considerably; or a nonlinear system itself tunes the 
noise level in 
order to enhance a particular chosen signal. Moreover, the noise level 
at which the response peaks depends on the frequency of the input signal. As a 
consequence, it may have potential applications in the detection of weak 
signals as well as in the selection of a signal of a particular frequency out 
of a host of signals of different 
frequencies\cite{Mantegna, Murali, Mohanty, Collins}. As a corollary, a 
biological system can tune noise level internally to select and enhance a 
desired signal\cite{Gluckman, Simonotto, Taba}. Apart from potential 
practical applications, it offers considerable theoretical challenges. The 
occurrence of SR in bistable systems\cite{Gamma, McN} has more or less been 
confirmed and reported fairly widely\cite{Gamma, Well}. However, its 
occurrence in periodic potentials is still debatable\cite{Kim}, though there 
have been some investigations in monostable\cite{Stocks} and periodic potential 
systems\cite{Dykman} as well.

There have been some discussion on SR in periodic potentials in view of obvious
practical importance of these potentials, such as in describing motion of
adatoms on crystal surfaces\cite{Graham}, superionic conductivity, RCSJ model 
of Josephson 
junction\cite{Kauf, Barone, Risken}, etc. The conventional SR is considered to 
occur in a bistable system at a temperature when the signal frequency matches 
the time rate of passage across the potential barrier of the nonlinear 
system\cite{Kim, McN}. Analogously, one would expect, for instance, the 
frequency dependent mobility, as a response, to peak as a function of 
temperature, at a signal frequency corresponding to the mean passage rate 
across a potential 
barrier of the periodic potential. However, instead, the mobility shows 
monotonic behaviour around that frequency. It does show a peak with 
temperature only at a much higher frequency. This mobility peaking with 
temperature is termed as a kind of dynamical resonance, unlike the SR 
satisfying the conventional criterion of frequency matching, and merely 
confined to intrawell motion\cite{Kim, Stocks, Dykman}. This aspect has been 
examined more closely in a recent numerical work\cite{PRE} on a sinusoidal 
potential considering hysteresis loop area as response to the input signal 
$F(t)=\Delta F\cos\omega t$.

Hysteresis loop area (HLA), in the average position-forcing, $(x-F)$, space, or
equivalently, the input energy lost by the system to the environment per period
of the external forcing $F(t)$, is considered as an appropriate measure of SR.
HLA has been considered as a quantifier of SR earlier too\cite{SRS, Iwa}. 
However, recently HLA was found to show SR behaviour close to what was shown 
by the amplitude of the average particle-position variable $x(t)$ in a bistable 
system\cite{Hein, Saikia,Sahoo,Jop}. HLA not only includes information of the 
mean amplitude $x_0$ of $x(t)$ but also its phase relationship with the 
external forcing or the input signal $F(t)$. Its significance as an appropriate 
quantifier of SR has also been pointed out recently\cite{Evstigneev,Jung}. 
Moreover, in a periodic potential, it is very likely that the particle forays 
into wells far away, on either direction, from the initial well. It makes the 
position variable unsuitable for any meaningful 
quantification of SR. It is found that the HLA or input energy loss takes 
account of motion whether it is in a single well or spread over several wells. 
Also, HLA shows qualitatively similar behaviour as the frequency dependent 
mobility calculated using the linear response theory\cite{Kim}. The HLA has 
recently been used as a quantifier of SR in a underdamped periodic potential 
system\cite{PRE}. 

In Ref.\cite{PRE} it was pointed  out that the peaking of HLA as a 
function of temperature at high frequencies ought not to be dismissed as 
merely a dynamical resonance\cite{Kim}. The phenomenon can be seen as a result 
of transitions between two dynamical states of the damped driven particle 
trajectories in the highly nonlinear (sinusoidal) potential. The two and only 
two dynamical states exist and are distinguished by their trajectory 
amplitudes and their phase relationship with the field $F(t)$. The relative 
stability of the two states changes as the temperature (noise strength) is 
varied.

It is hard to prove analytically that an underdamped particle moving in a 
sinusoidal potential driven by a periodic force of frequency close to the 
natural frequency at the bottom of a well of the potential and subjected to 
fluctuating forces can have only two dynamical states of its trajectories as 
it was found numerically. Solution of only the simplest situation of the 
deterministic motion of a driven pendulum without friction,\\
$m\frac{d^2x}{dt^2}=-\frac{\partial}{\partial x}(-\sin x)+F_0\cos\omega t$,\\
given the initial condition $(x(0),v(0))$, can be obtained analytically and its 
phase trajectories drawn\cite{Laksh}. Notice that the periodic potentials 
$U_0(x)=-\sin x$ and $U_0(x)=\cos x$ are equivalent except for a phase 
difference of $\pm \frac{\pi}{2}$. Also, the trajectories with initial 
conditions $(x(0),0)$ and $(2\pi-x(0),0)$ should be identical except that they 
are in opposite directions in case of closed orbits and in $\pm v$ sectors in 
case of travelling solutions. In other words, the two trajectories should have 
the same amplitude but different phase relationship with the external forcing.
However, for different initial conditions ($x(0),v(0)$), they can have
different trajectories and not necessarily confined to just two particular 
trajectories. But, when damping is present, it is easy to appreciate that if 
two trajectories have different phases they will necessarily have different 
amplitudes and hence different energy contents. This can be seen as follows.

\begin{figure}[htp]
\centering
\includegraphics[width=10cm,height=8cm,angle=0]{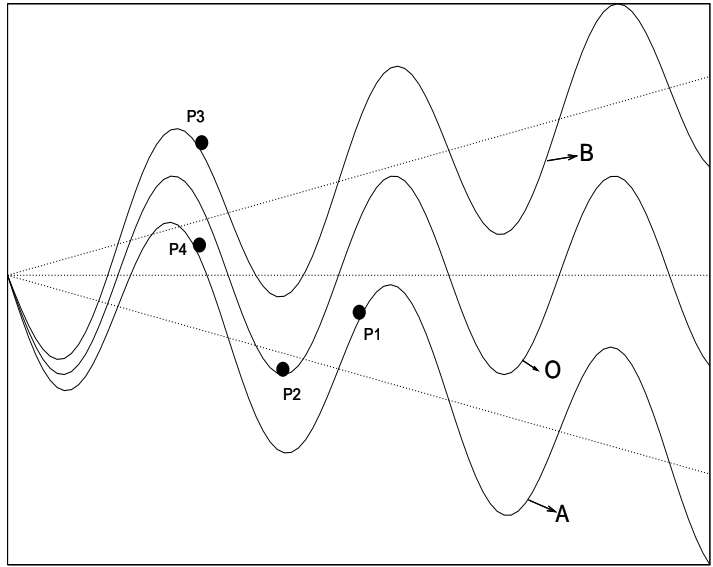}
\caption{The figure shows the periodic potential when $F(t)=0$ (curve $O$),
$F(t)=\Delta F$ (curve $A$) and $F(t)=-\Delta F$ (curve $B$).}
\label{fig:edge}
\end{figure}

Consider the two extreme cases of particle motion being (i) in-phase, and (ii)
completely out-of-phase with the external drive. Fig. 1 shows the periodic 
potential when $F(t)=0$ (curve O), $F(t)=\Delta F$ (curve A), and 
$F(t)=-\Delta F$ (curve B), where $F(t)=\Delta F\cos \omega t$, so that the 
total potential $V(x)=-\sin x-x\Delta F\cos \omega t$. Consider the extreme 
position P1 of the particle on the curve A when $F(t)=\Delta F$. In the next 
moment the $F(t)$ will decrease and also the particle position moves to the 
left so that $F(t)$ and $x(t)$ are in phase. They being in phase implies that 
when $F(t)=0$ the particle is at the potential bottom P2 and when 
$F(t)=-\Delta F$ the particle is at P3, the extreme position on the left and 
is about to roll down the potential hill as $F(t)$ begins to rise from 
$-\Delta F$. From the figure, therefore, one could notice that in this case 
of in-phase situation the particle always moves on the stiffer slope of 
the potential $V(x)$. On the other hand, if we were to begin from the particle 
position P4 on A while $F(t)=\Delta F$, the particle always moves lying on the 
gentler slope of $V(x)$ in this completely out-of-phase case. Since the force 
experienced by the particle due to the potential in the two cases have 
different values one would naturally expect the damped particle to have 
different amplitudes of motion and hence have different energy losses due to 
frictional forces. The two dynamical states are thus distinct.

At low temperatures, for given $F(t)$, depending only on the initial 
conditions the system chooses one of the two dynamical states. These two states
are quite stable\cite{PRE} and no transition 
occurs between them. However, as the temperature is increased transition takes 
place between the dynamical states and, as a result, the relative population of 
these states changes. Thus, the mean amplitude of the trajectories and the 
overall phase when averaged over the initial conditions vary with temperature. 
The hysteresis loop area shows a peak at a temperature indicating stochastic 
resonance where the transition rate between the dynamical states acquires a 
particular value. It is also important to notice that well before the HLA 
peaks, the particle begins to surmount the potential barrier of $V(x)$ and at 
resonance the motion is no longer confined to a single well of $V(x)$; the 
inter-well transitions become quite numerous. Of course, the inter-well 
transition rate is still quite low compared to the transition rate between the 
dynamical states\cite{PRE}.

In the present work, we show that a driven underdamped particle exhibits 
stochastic resonance in (i) a periodic bistable potential, 
$U(x)=\frac{2}{3}(\cos x +\cos 2x)$, and also in (ii) washboard potentials 
(tilted sinusoidal as well as tilted periodic bistable potential). In this
work we use HLA as a quantifier of SR but also supported by the signal-to-noise 
ratio (SNR).

The appropriateness of the HLA as a quantifier of SR vis-a-vis the SNR has 
been discussed extensively in 
Ref.\cite{Evstigneev}. However, the SNR has been used in many earlier works 
to discuss SR. Without going into the relative merits of these quantifiers we 
present the SNR to complement extensive results that we show in terms of 
HLA. As has been pointed out\cite{Evstigneev} the quantifiers do not peak at 
the same temperature. However, our main contention that SR is a distinct 
possibility in periodic potentials is reinforced by the results of SNR.
\begin{figure}[htp]
\centering
\includegraphics[width=10cm,height=8cm,angle=0]{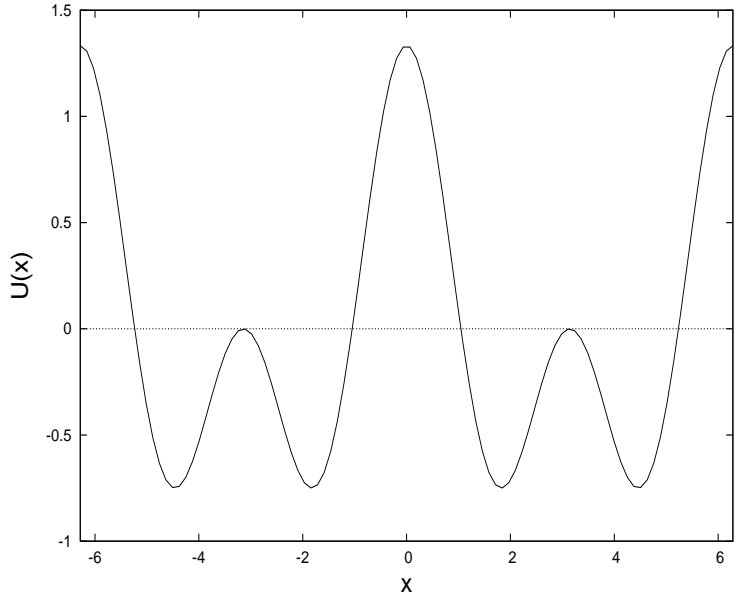}
\caption{The periodic bistable potential $U(x)=\frac{2}{3}(\cos x +\cos 2x)$ 
is shown with two similar subwells in a well of the potential.}
\label{fig:edge}
\end{figure}

In the case of bistable periodic potential too, effectively, two and only two 
dynamical states (one in-phase and one out-of-phase) of trajectories of a 
periodically driven underdamped particle are obtained. Since in a potential 
well of $U(x)$ there exist two similar subwells (Fig. 2) there are two (one 
in-phase and one out-of-phase) states in each subwell. The in-phase 
(out-of-phase) 
state in one subwell has the same amplitude and phase relationship as the 
in-phase (out-of-phase) state in the other subwell. Therefore, energetically 
there exist only one in-phase and one out-of-phase states of trajectories. 
However, as time progresses, these trajectories can also be in wells and 
subwells other than the initial ones, the basins of attraction of the dynamical
states can be identified in different wells and subwells. Consequently, the 
basins of attraction of the dynamical states become more complex than in case 
of sinusoidal potentials where only trajectories in different wells are 
distinguished. 

In the case of periodic bistable potentials HLA, as also the signal-to-noise 
ratio, shows a peaking behaviour, just as in the case of sinusoidal potential, 
as the temperature is varied. The non-monotonic behaviour of HLA as a function 
of temperature is related to the relative stability of the in-phase 
and out-of-phase states of trajectories. Moreover, the particle does make 
inter-well transitions also.

The washboard potentials, that is a slanted sinusoidal potential and slanted
bistable periodic potential, naturally yield finite particle drifts. As a
consequence the hysteresis loops do not close. However, upon correcting for 
the drift factor, as explained in Sec. III, the hysteresis loops close and the 
closed hysteresis loop areas could be made to agree with the input energy loss.
The HLA (or the input energy loss) so obtained again shows peaking behaviour 
as the temperature is varied. This is a clear signature of SR in the washboard 
potential. The occurrence of SR could again be explained in terms of transition
between the two dynamical (one in-phase and one out-of-phase) states of 
particle trajectories that are realized even in these washboard potentials. 

\section{The model}

We consider motion of an underdamped particle along (i) a periodic bistable potential $U(x)=\frac{2}{3}V_0(\cos x +\cos 2x)$, (ii) a tilted sinusoidal potential
$U_1(x)=-V_0\sin kx-F_0x$, and (iii) a tilted bistable potential 
$U_2(x)=U(x)-F_0x$, 
where the tilt $F_0$ represents a constant force. As mentioned earlier (Fig.2)
$U(x)$ has two subwells of equal depth (bistable) in each periodic well of the 
potential. $U(x)$ is symmetric about $x=n\pi$, where, $n=\mp 1,\mp 2, \cdots$. 
The latter two potentials we shall refer to as the washboard potential and the
bistable washboard potential, respectively, for $F_0\neq 0$ but small. In the 
bistable washboard potential $U_2(x)$ the two subwells exist but are no longer
identical as they were in $U(x)$.

A particle of mass $m$ moving in a medium of friction coefficient $\gamma$
along a potential $V(x)$ and driven by an external periodic forcing $F(t)=
\Delta F \cos \omega t$ and subjected to a Gaussian white noise $\xi (t)$, is 
described here by the Langevin equation,
\begin{equation}
m\frac{d^{2}x}{dt^{2}}=-\gamma\frac{dx}{dt}-\frac{\partial{V(x)}}{\partial
x}+F(t)+\sqrt{\gamma T}\xi(t).
\end{equation}
The fluctuating forces $\xi (t)$ satisfy: $<\xi (t)>=0$ and 
$<\xi(t)\xi(t^{'})>=2\delta(t-t^{'})$. The temperature $T$ is in units of the 
Boltzmann constant $k_B$. We take $V(x)=U(x), U_1(x)$, or $U_2(x)$ separately 
to discuss the nature of particle motion.

The equation is written in dimensionless units by setting $m=1$, $V_0=1$, 
$k=1$. The Langevin equation, with reduced variables denoted again now by the
same symbols, corresponding to Eq. (2.1) is written as

\begin{equation}
\frac{d^{2}x}{dt^{2}}=-\gamma\frac{dx}{dt}
+cos x +F(t)+\sqrt{\gamma T}\xi(t).
\end{equation}
 
The noise variable, in the same symbol $\xi$, satisfies exactly similar 
statistics as earlier.

\section{Numerical Results}

We adopt the same numerical procedures as described in Ref.\cite{PRE}. The
drive (signal) frequency ($\omega=\frac{2\pi}{\tau}$) is chosen to be close 
to but a little smaller than the natural frquency at the bottom of the wells 
of the potentials. However, they are not exactly equal to the respective 
natural frequencies (without damping). For the periodic bistable potential 
$U(x)$, we take the period $\tau$ equal to 4.8. The same optimum $\tau$ is 
taken also for the bistable washboard potential $U_2(x)$. Similarly, we 
continue with $\tau=8$ as in Ref.\cite{PRE} for the washboard potential 
$U_1(x)$.
 
With the periods $\tau$ of the external forcing $F(t)=\Delta F\cos\omega t$, 
we obtain the trajectories $x(t)$ for given initial conditions $(x(0),v(0))$ 
numerically\cite{Nume, SRS} by solving the Langevin equation (2.2) and 
calculate the input energy, or work done by the field on the system, $W$, in 
a period $\tau$, as\cite{Seki}:
\begin{equation}
W(t_0,t_0+\tau)=\int_{t_0}^{t_0+\tau}\frac{\partial U_e(x(t),t)}{\partial t}dt,
\end{equation}
where, the effective potential $U_e(x(t),t)=V(x)-xF(t)$, and $V(x)=U(x), 
U_1(x)$, or $U_2(x)$ as applicable. 
Therefore,
\begin{equation}
W(t_0,t_0+\tau)=-\int_{F(t_0)}^{F(t_0+\tau)}xdF=A,
\end{equation}
where $A$ is the magnitude of the HLA. The average input energy per period, 
$\overline{W}$, averaged over an entire trajectory spanning $N_1$ periods of
$F(t)$, is
\begin{equation}
\overline{W}= \frac{1}{N_1}\sum_{n=0}^{n=N_1}W(n\tau,(n+1)\tau)=\overline{A}.
\end{equation}
Typically, $N_1$, ranges between $10^5$ to $10^7$, as required. 

At very low temperatures $\overline{W}$ depends very strongly on the initial 
conditions ($x(0),v(0)$). This is because whether the trajectories are in the 
in-phase state (with a small phase difference $\phi_1$ between $x(t)$ and 
$F(t)$) or in the out-of-phase state (with a large phase difference $\phi_2$) 
is determined by the initial condition. And, $\overline{W}$ depends on the kind
of trajectory. Moreover, at such temperatures the particle remains trapped in 
the initial well of the potential and also no transition between in-phase and 
out-of-phase states of trajectories takes place in a well. However, as the 
temperature is gradually increased the intra-well transitions between the 
in-phase and out-of-phase states begin to take place and subsequently, 
inter-well transitions also become frequent. Therefore, as the temperature is 
increased the dependence of $\overline{W}$ on the initial conditions 
($x(0),v(0)$) weakens. However, it is always sensible to ensemble average 
$\overline{W}$ over all possible initial conditions and obtain the average 
input energy per period $<\overline{W}>$, which is also equal to the mean 
hysteresis loop area $<\overline{A}>$, Eq. (3.3).

As mentioned earlier, the response $x(t)$ in the two dynamical phases of 
trajectories not only have different phase relationship with the forcing 
$F(t)$ but the mean amplitude $x_0$ of $x(t)$ in the two cases are very 
different. Therefore, apart from the inherent stochastic nature of $W$ (
Eq. 3.1) the values of $W$ are different depending on whether, during the 
particular period (of $F(t)$) in question, the system is in the in-phase or 
in the out-of-phase state of trajectory or makes transition(s) between the 
states. A clear picture is revealed by the distribution $P(W)$ of input 
energies at various temperatures. For sinusoidal potentials the evolution of 
$P(W)$ with temperature furnishes important information on the occurrence of 
SR\cite{Saikia, Sahoo, Jop}. Also, since the entire stretch of the 
trajectories consists of a mixture of in-phase (with phase $\phi_1$) and 
out-of-phase (with phase $\phi_2$) states, the average phase lag $\phi$ 
must lie between $\phi_1$ and $\phi_2$. Naturally, $\phi$ is a function of
temperature\cite{LGamma, Dykman1}. The phase $\phi$ is obtained 
numerically\cite{PRE} by calculating the mean hysteresis loops 
$<\overline{x}(F(t_i))>$, where,
\begin{equation}
\overline{x}(F(t_i))=\frac{1}{N_1}\sum_{n=0}^{n=N_1}x(F(n\tau+t_i)),
\end{equation}
for all $[0\leq t_i<\tau]$. 

In the following subsections we present and discuss our numerical results 
separately for the three cases of underdamped particle motion in potentials 
$U(x), U_1(x)$, and $U_2(x)$ driven by field $F(t)$ at various temperatures 
$T$. We take the dimensionless friction coefficient $\gamma=0.12$ and initial 
velocity $v(0)=v(t=0)=0$ for all cases. The initial position $x(0)=x(t=0)$ are
chosen at 99 equispaced points $x_i, i=1,2,\cdots,99$ between the two 
consecutive peaks, e.g., [$0<x_i<2\pi$], for the potential $U(x)$. Unless 
otherwise explicitely stated the amplitude $\Delta F$ of $F(t)$ is taken equal 
to 0.2 and the tilt $F_0=0.1$. 

In the case of washboard potential $U_1(x)$ and the bistable washboard 
potential $U_2(x)$, with constant tilt $F_0$ the particle does acquire a mean 
drift (of velocity $\overline{v}(F_0)$) at elevated temperatures and 
consequently the hysteresis loops do not close. Therefore, a correction is 
required to make the loops close. In a period, $F(t)$ changes from 
$F_0-\Delta F$ to $F_0+\Delta F$ and then back to $F_0-\Delta F$ in a 
cosinusoidal manner. During a period the particle moves on the average a 
distance of $\tau\overline{v}(F_0)$. Approximating the variation of $F(t)$
to be linear the mean work done during a period $\tau$ as a result of the
mean particle displacement is equal to 
$\frac{1}{2}\times 2\times\Delta F\times\tau\overline{v}(F_0)$. Therefore, the 
required correction to the hysteresis loop area $<\overline A>$ due to the 
mean drift equals $\Delta F\tau\overline{v}(F_0)$. This simple approximate 
correction closes the hysteresis loops thereby enabling us to calculate the 
HLA and equivalently the $<\overline{W}>$.
 
\subsection{Dynamical states of trajectories}

\begin{figure}[htp]
  \centering
\includegraphics[width=10cm,height=7cm,angle=0]{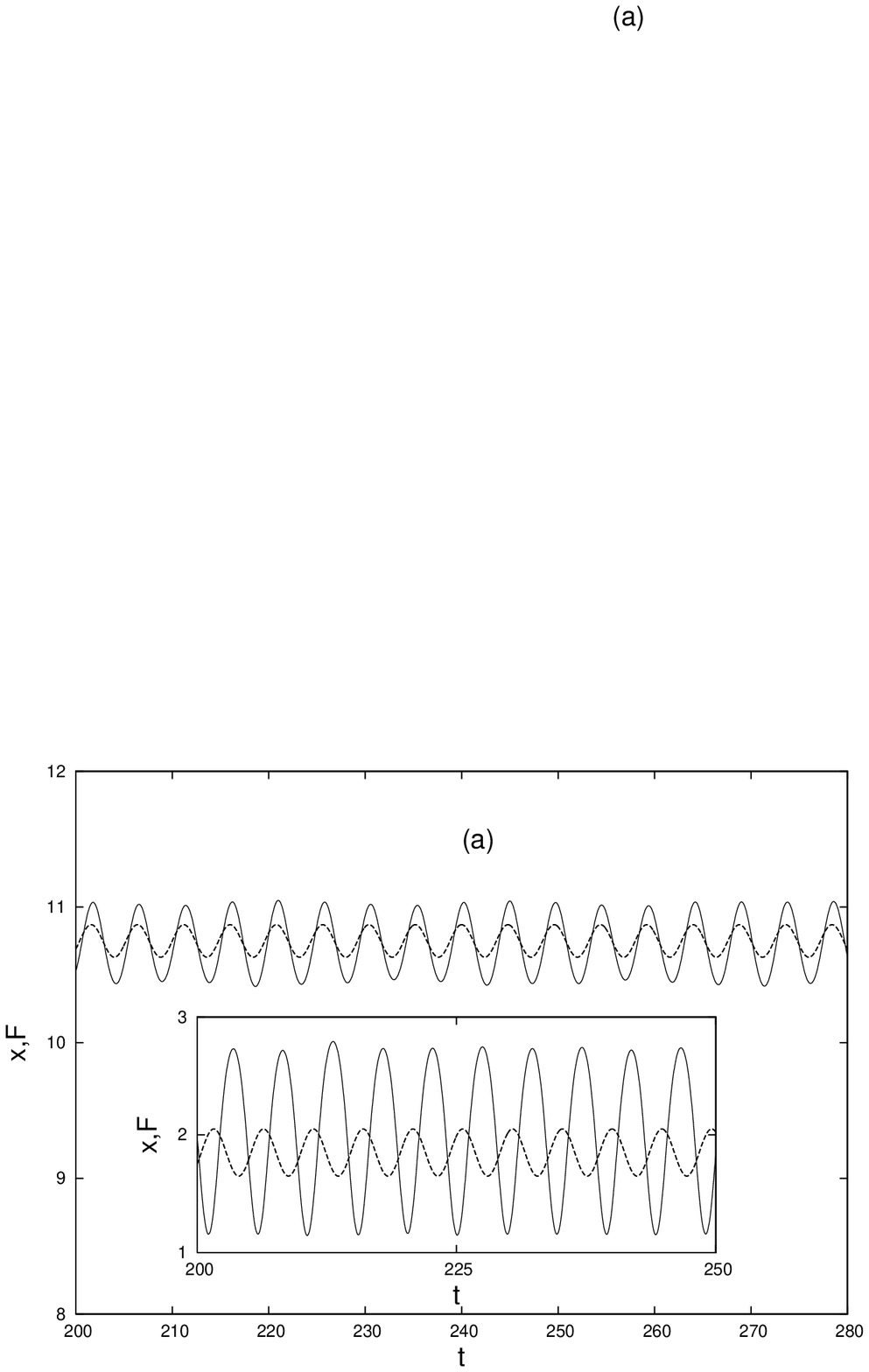}
\vspace{0.4cm}
\includegraphics[width=10cm,height=7cm,angle=0]{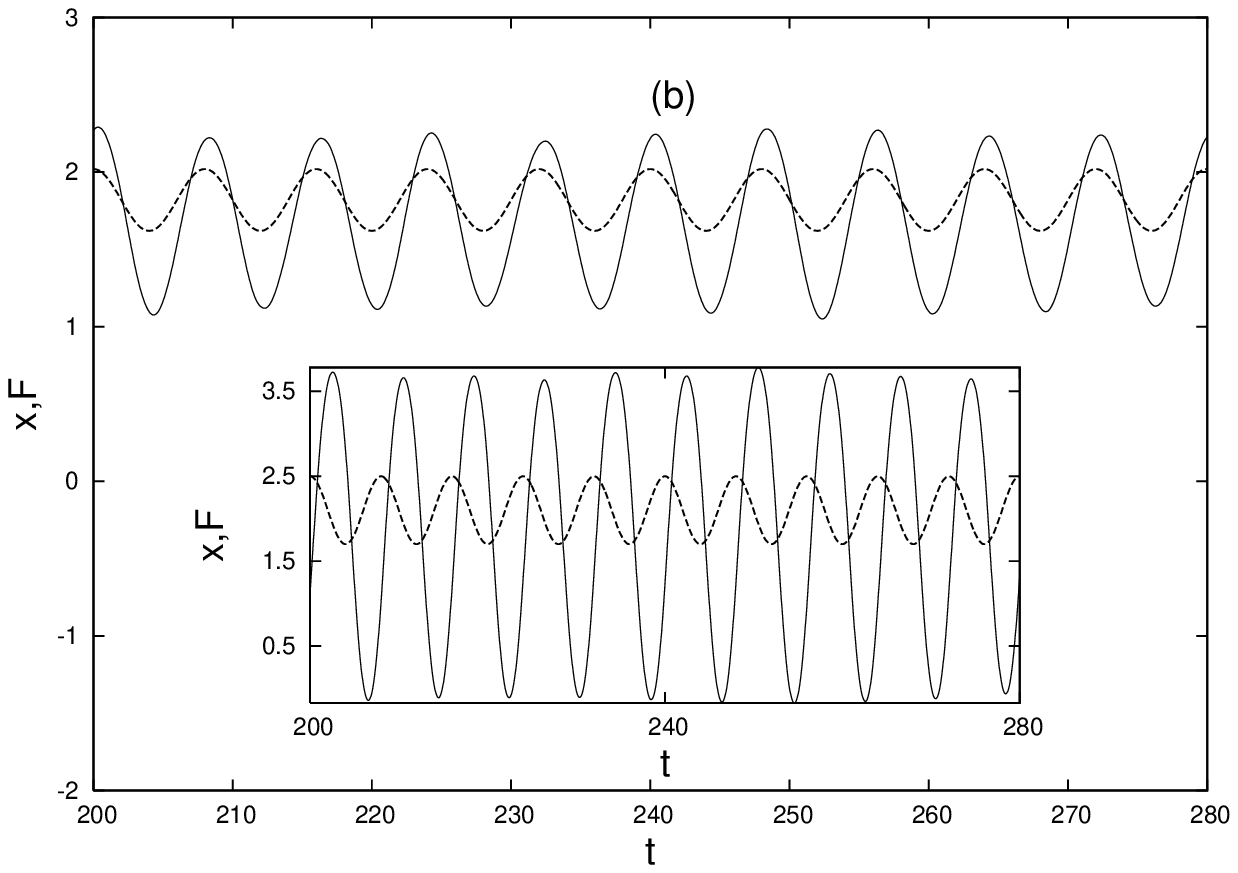}
\caption{Plot of particle trajectories $x(t)$
(a) for $U(x)$ at $T=0.0005$, and (b) for $U_1(x)$ at $T=0.001$. $F(t)$ 
(dashed line) is also included for comparison. The phase lags
$\phi_1\sim 0.08\pi$ and $\phi_2\sim 0.8\pi$ (inset) for $U(x)$ and 
$\phi_1\sim 0.08\pi$ and $\phi_2\sim 0.6\pi$ (inset) for $U_1(x)$.}
\end{figure}

\begin{figure}[htp]
 \centering
\includegraphics[width=15cm,height=10cm,angle=0]{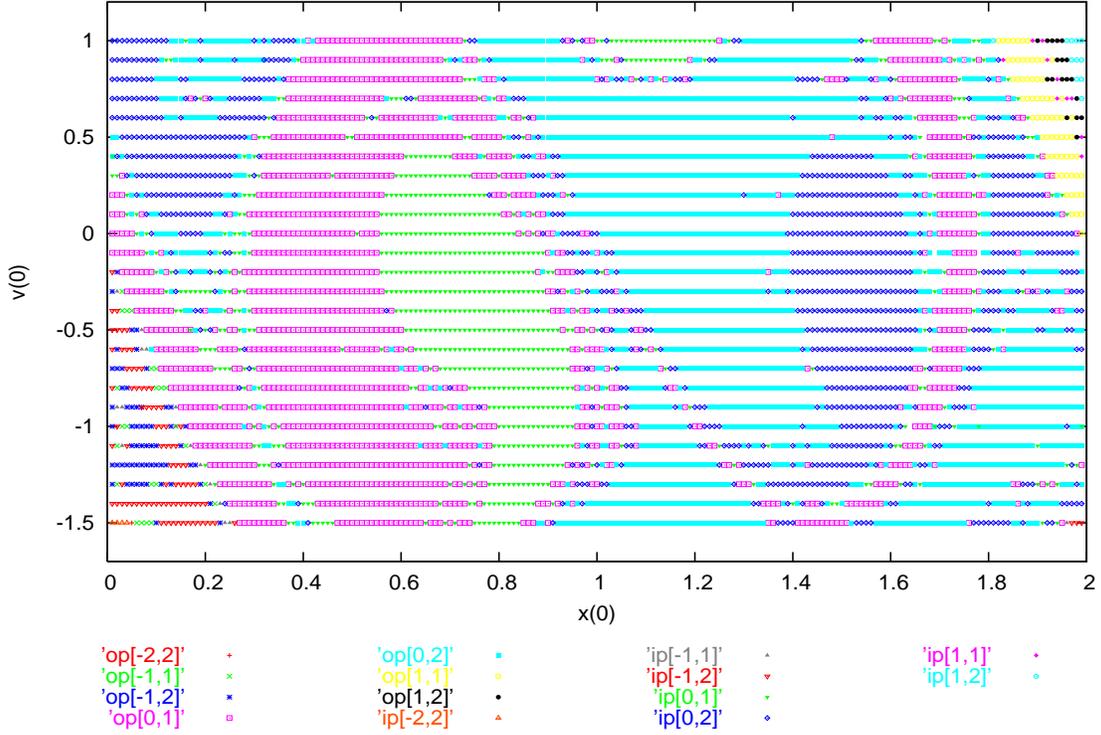}
\caption{The figure shows the basins of attraction for
bistable potential at $T=0.0005$. The dynamical states of the
particle are represented by ip- (in-phase) and op- (out-of-phase)
states with respect to applied periodic force. The wells and subwells in which
the dynamical states exist are represented by the indices $l$ and $m$ within
braces; e.g., op[0,1] indicates that the out-of-phase trajectory of the 
particle is in the (left) subwell-1 of the (initial) zeroth well of the 
potential.}
\label{fig:edge}
\end{figure}

When the underdamped particle moves along the potentials 
$U(x) (=\frac{2}{3}(\cos x+\cos 2x))$ and $U_1(x) (=-\sin x-F_0x)$
and driven by the external periodic field $F(t)$ the trajectories are 
essentially of just two kinds (states). The in-phase states correspond to 
trjectories $x(t)$ which lag behind $F(t)$ by a small phase $\phi_1$, whereas
the out-of-phase states which lag behind $F(t)$ by a large phase $\phi_2$. Of
course, $\phi_1$ and $\phi_2$ are approximate average values which weakly 
depend on the potential and the temperature. At low temperatures $\phi_1$ and 
$\phi_2$ essentially maintain the same values at each period of $F(t)$, Fig.3.
The potential $U(x)$ has two similar subwells (Fig. 2) in a well, the
trajectories in either subwell are just the same two states. Here, these 
in-phase and out-of-phase states are identified by the symbols ip($l,m$) and 
op($l,m$), respectively. $l$ indicates well number, for example, $l=0$ for the 
initial well, $l=-1(+1)$ for the first well to the left(right) of the initial 
well. $m$ indicates the subwell number: 1(2) for the left(right) subwell. These 
states are truly dynamical states. The basins of attraction of these states 
for the potential $U(x)$ is given in Fig. 4, as an illustration at temperature 
$T=.0005$.

\begin{figure}[htp]
  \centering
\includegraphics[width=9cm,height=6cm,angle=0]{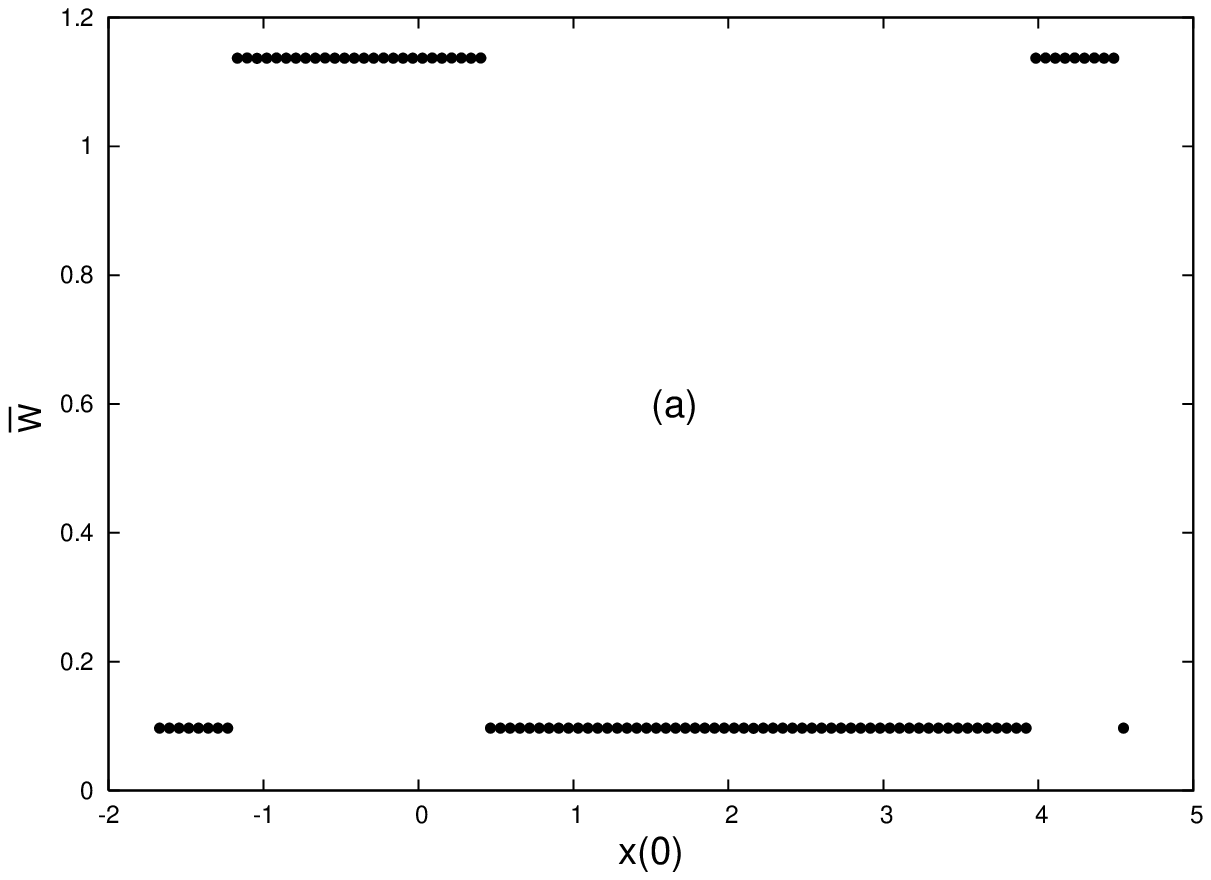}
\vspace{0.4cm}
\includegraphics[width=9cm,height=6cm,angle=0]{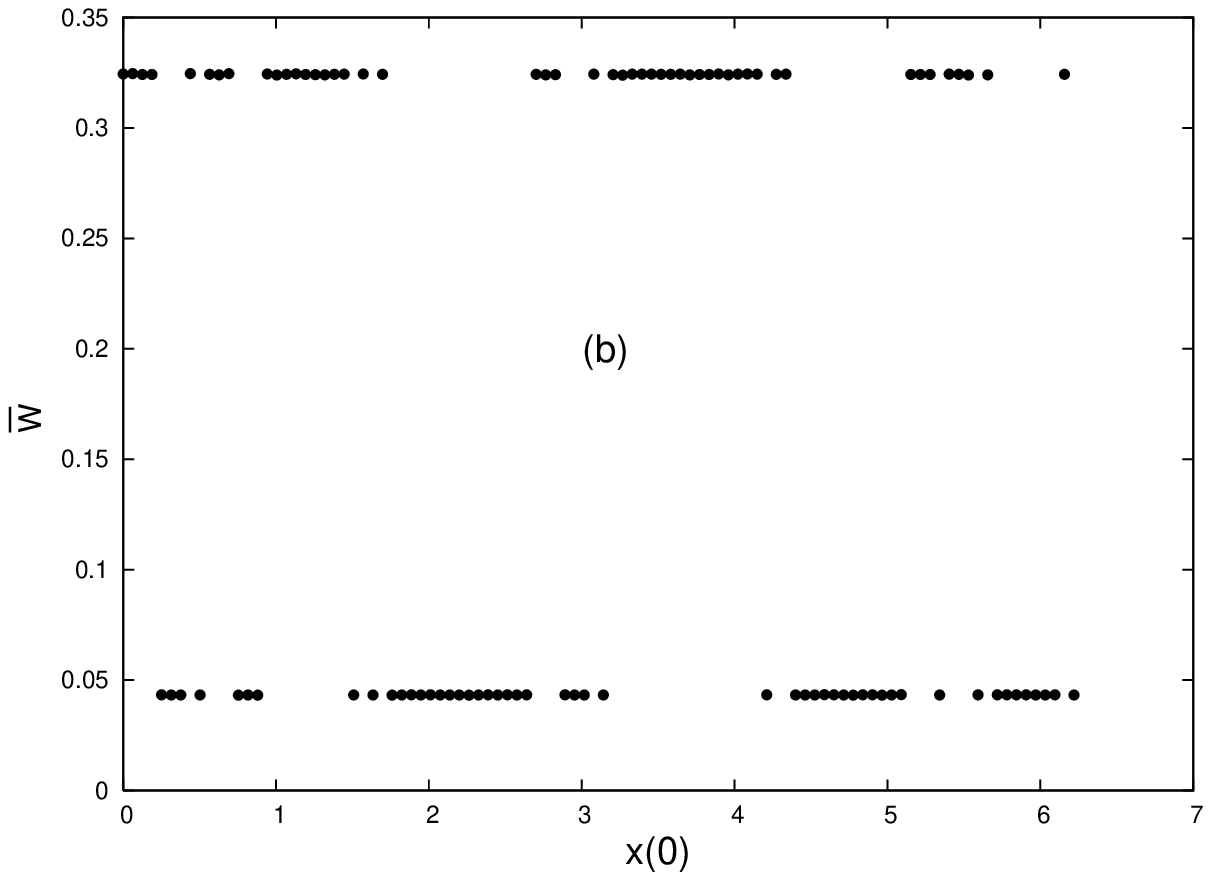}
\caption{Plot of $\overline{W}$ with $x(0)$ for $U_1(x)$ at $T=0.001$ (a),
and $U(x)$ at $T=0.0005$ (b).}
\end{figure}

\begin{figure}[htp]
  \centering
\includegraphics[width=9cm,height=6cm,angle=0]{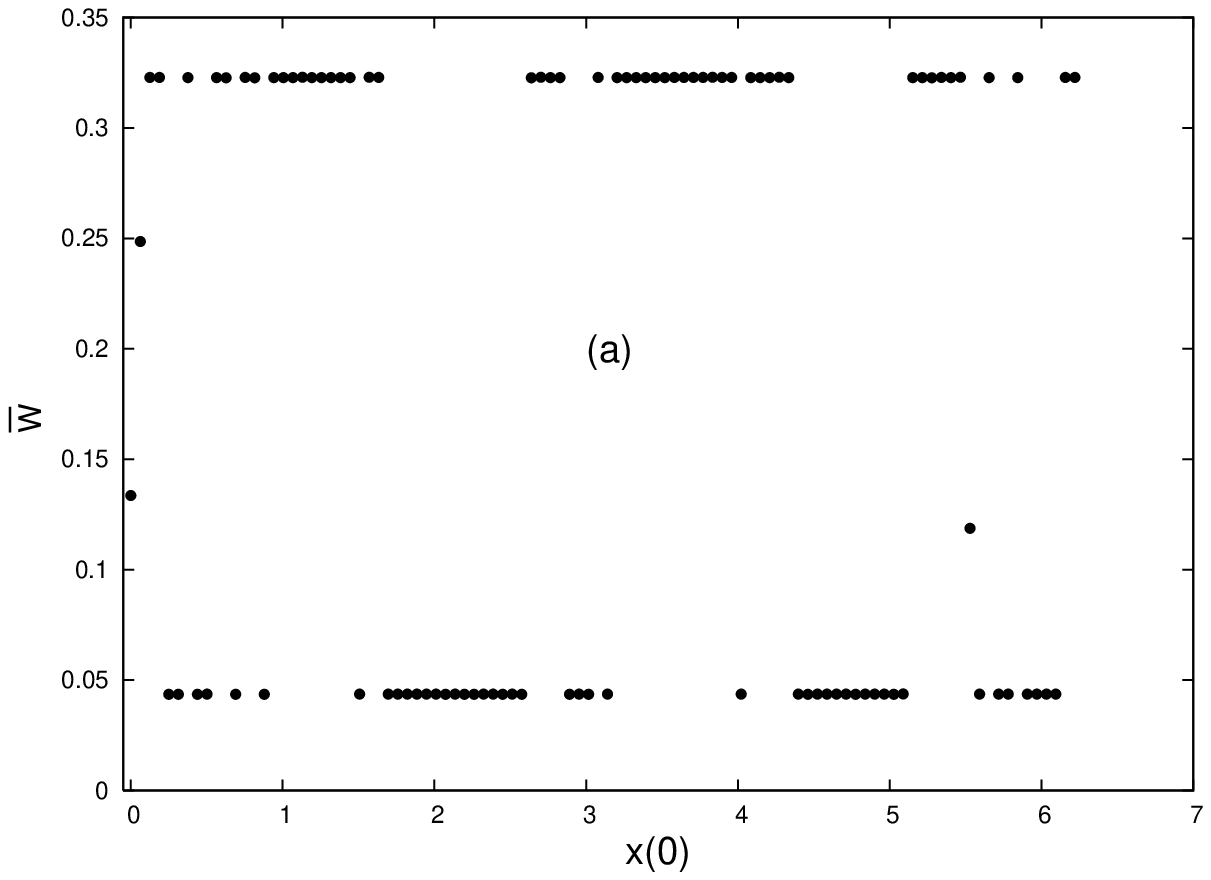}
\vspace{0.4cm}
\includegraphics[width=9cm,height=6cm,angle=0]{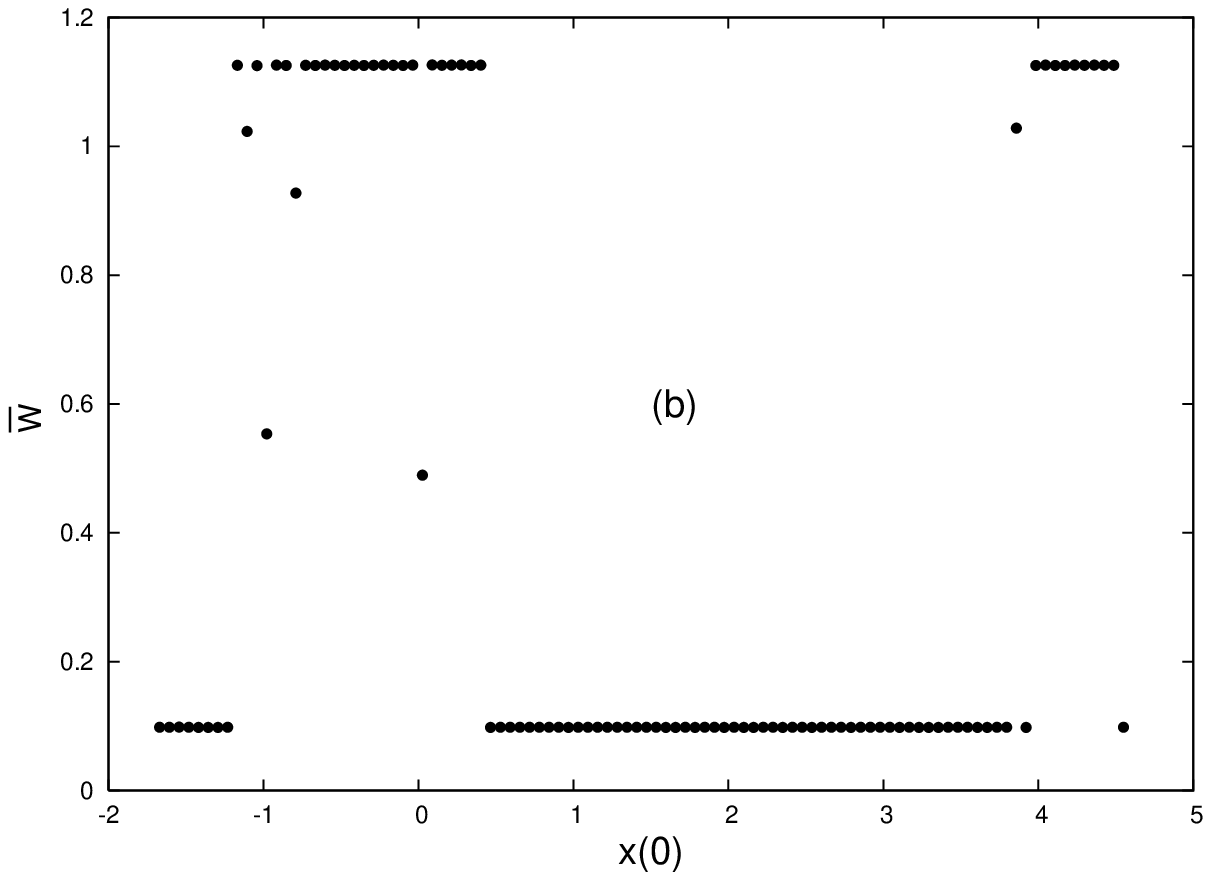}
\caption{Plot of $\overline{W}$ with $x(0)$ for $U(x)$ at $T=0.001$ (a),
$U_1(x)$ at $T=0.004$ (b).}
\end{figure}

At temperature $T=.0005$ the two states are quite stable and no transition 
between them could be observed. Though the potential barrier between any two
consecutive wells have almost the same value for $U(x)$ and $U_1(x)$, their 
well bottoms are quite dissimilar and hence the input energy per period 
$\overline{W}$ are very different, Fig. 5. For $U_1(x)$, $\overline{W}$ are
about 0.1 (in-phase) and 1.15 (out-of-phase), whereas for $U(x)$ they are
about 0.04 (in-phase) and 0.32 (out-of-phase). Naturally, transitions between 
the two states for $U(x)$ occur at lower temperature than in case of $U_1(x)$,
Fig. 6.

In the case of $U(x)$ by the temperature $T=0.001$ the indications of 
out-of-phase state going to the in-phase state, in a subwell, could be found and
by $T=0.0015$ a substantial fraction of the out-of-phase states have jumped to 
the in-phase state. At $T=0.002$ all the out-of-phase states have made way to 
the in-phase state due to thermal fluctuations. Therefore, close to $T=0.002$ 
the system have the lowest average input energy $<\overline{W}>$ (or 
$<\overline{A}>$) per period of $F(t)$. And, from upward of $T=0.003$, the 
in-phase states begin to jump to the out-of-phase state. At $T=0.005$ even 
inter-subwell transitions could also be observed. The corresponding 
temperatures for the case of $U_1(x)$ are much higher.

\begin{figure}[htp]
  \centering
\includegraphics[width=9cm,height=6cm,angle=0]{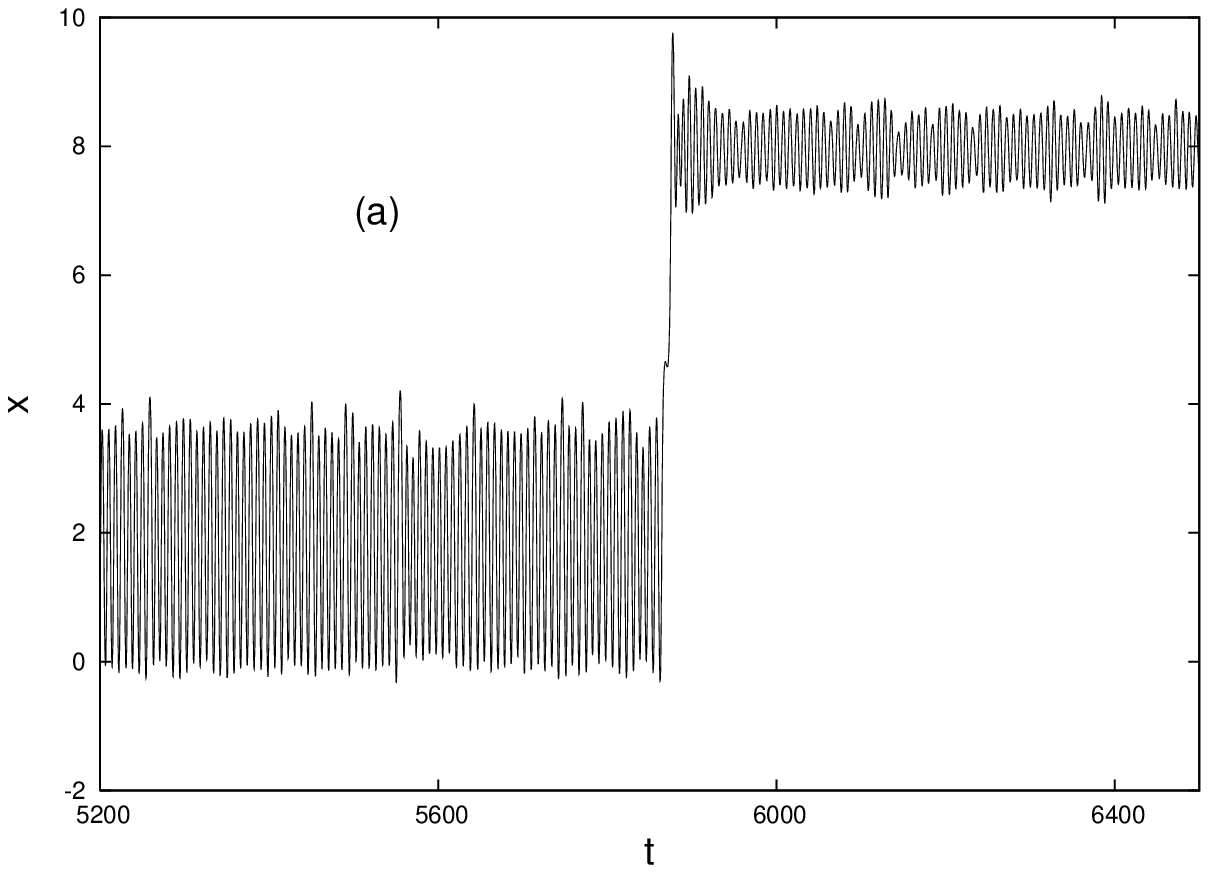}
\vspace{0.4cm}
\includegraphics[width=9cm,height=6cm,angle=0]{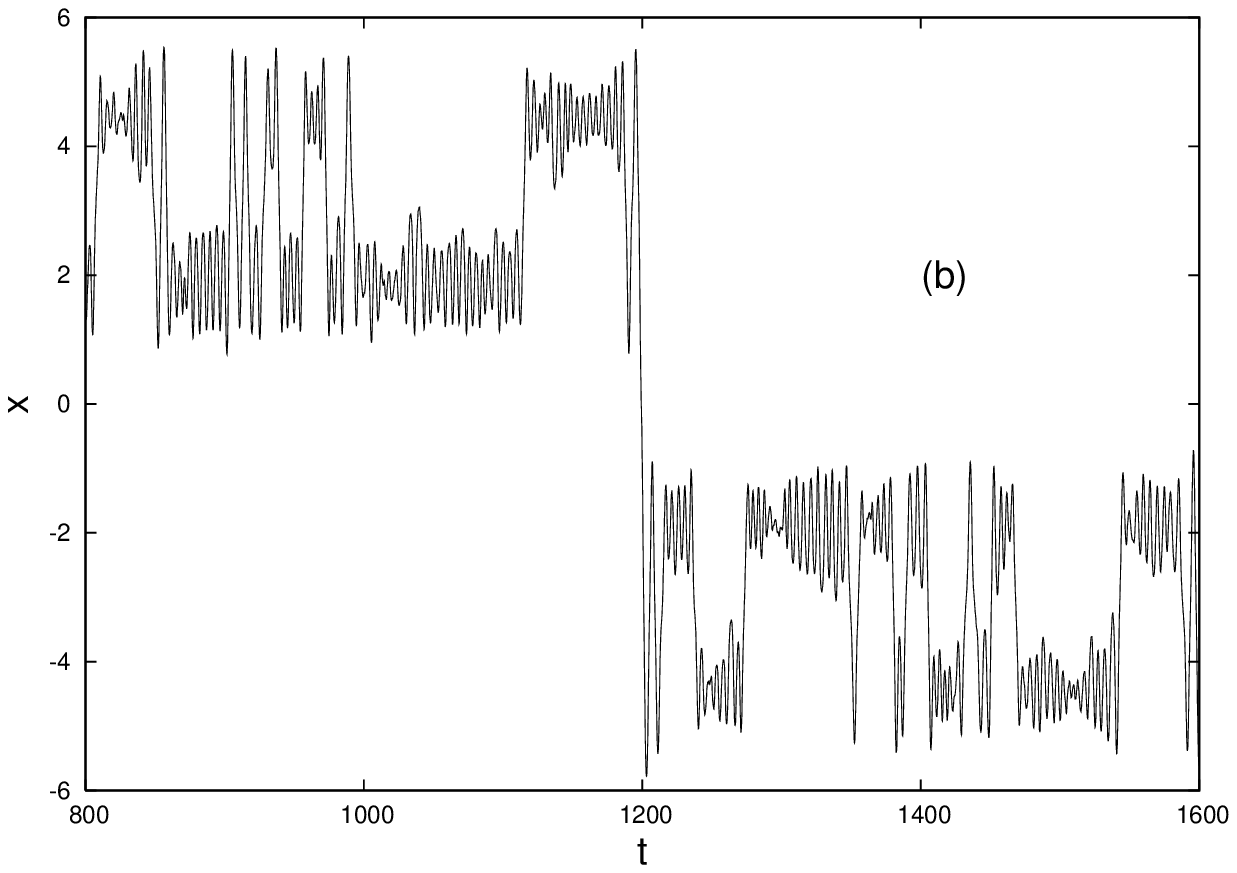}
\caption{Plot of $x(t)$ shows an inter-well transition for $U_1(x)$ at $T=0.01$ (a). A similar transition is seen for $U(x)$ at $T=0.23$ (b).}
\end{figure}

For the system with potential $U_1(x)$ the out-of-phase states begin going over 
to the in-phase state at around $T=0.004$ and the process completes at about
$T=0.006$. The transition from the all-in-phase state to the out-of-phase state
begins at about $T=0.009$. Interestingly, however, the inter-well transitions
for $U_1(x)$ takes place at a much lower temperature (about $T=0.01$) than in 
case of the system with potential $U(x)$ which is only at about $T=0.23$, 
Fig. 7. 

As stated earlier the periodic bistable potential $U(x)$ have essentially
two states because the two subwells are energetically identical. In the case
of the bistable washboard potential $U_2(x) (=U(x)-F_0x)$ the two subwells 
become dissimilar
and the right subwell is energetically lower than the left one and we get two
states corresponding to each subwell. However, the in-phase state in the left
subwell becomes unstable for amplitude $\Delta F>0.19$ of the drive $F(t)$.
Therefore, in our case, we have only one in-phase state to consider in the 
left subwell.

In case of $U_2(x)$, $\overline{W}$ is about 0.27 for the out-of-phase state
in the left subwell (henceforth called subwell-1) and $\overline{W}$ for the 
two states in the other subwell (subwell-2) are about 0.03 (in-phase) and 0.37 
(out-of-phase) at $T=0.0005$. By the temperature $T=0.001$ most of the 
(out-of-phase) states in subwell-1 go over to the states in subwell-2 and by 
$T=0.0015$ no states in subwell-1 survives. As the temperature is gradually 
increased the out-of-phase states in subwell-2 begin to jump over to the 
in-phase state in the same subwell and by $T=0.003$ we only have in-phase 
states in subwell-2 and $<\overline{W}>$ acquires a minimum value. By $T=0.009$
the particles begin to leave the in-phase state for the out-of-phase state in 
subwell-2 and we get a mixture of the two states in the same subwell. As the
temperature is increased further, the transitions back and forth to subwell-1
begin at about $T=0.015$. Inter-well transitions begin, at a very slow rate,
at about $T=0.08$, which is much higher than the corresponding temperature for
the washboard potential $U_1(x)$ but lower than that for $U(x)$. The presence 
of two subwells in $U(x)$ and $U_2(x)$ makes the effective bottom of the wells
flatter (smaller curvature) than in case of $U_1(x)$ and hence the inter-well
transition rates smaller (see Eq. (3.5) below).

At $T>0.003$ ($T>0.01$) transitions from in-phase states to the out-of-phase 
states increase rapidly for $U(x)$ ($U_1(x)$), and consequently so does the 
$<\overline{W}>$, leading ultimately to the SR condition. Similar is the
situation for $U_2(x)$.
 
\subsection{Hysteresis loss and stochastic resonance}

\begin{figure}[htp]
\centering
\includegraphics[width=10cm,height=8cm,angle=0]{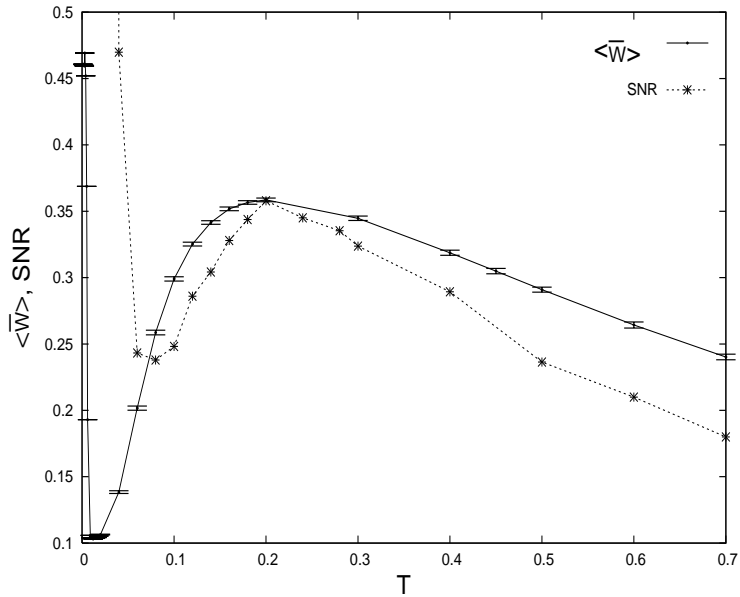}
\caption{Plot of $\langle \overline{W} \rangle$ and SNR as a function of $T$ 
for the washboard potential $U_1(x)$.}
\label{fig:edge}
\end{figure}

\begin{figure}[htp]
  \centering
\includegraphics[width=10cm,height=8cm,angle=0]{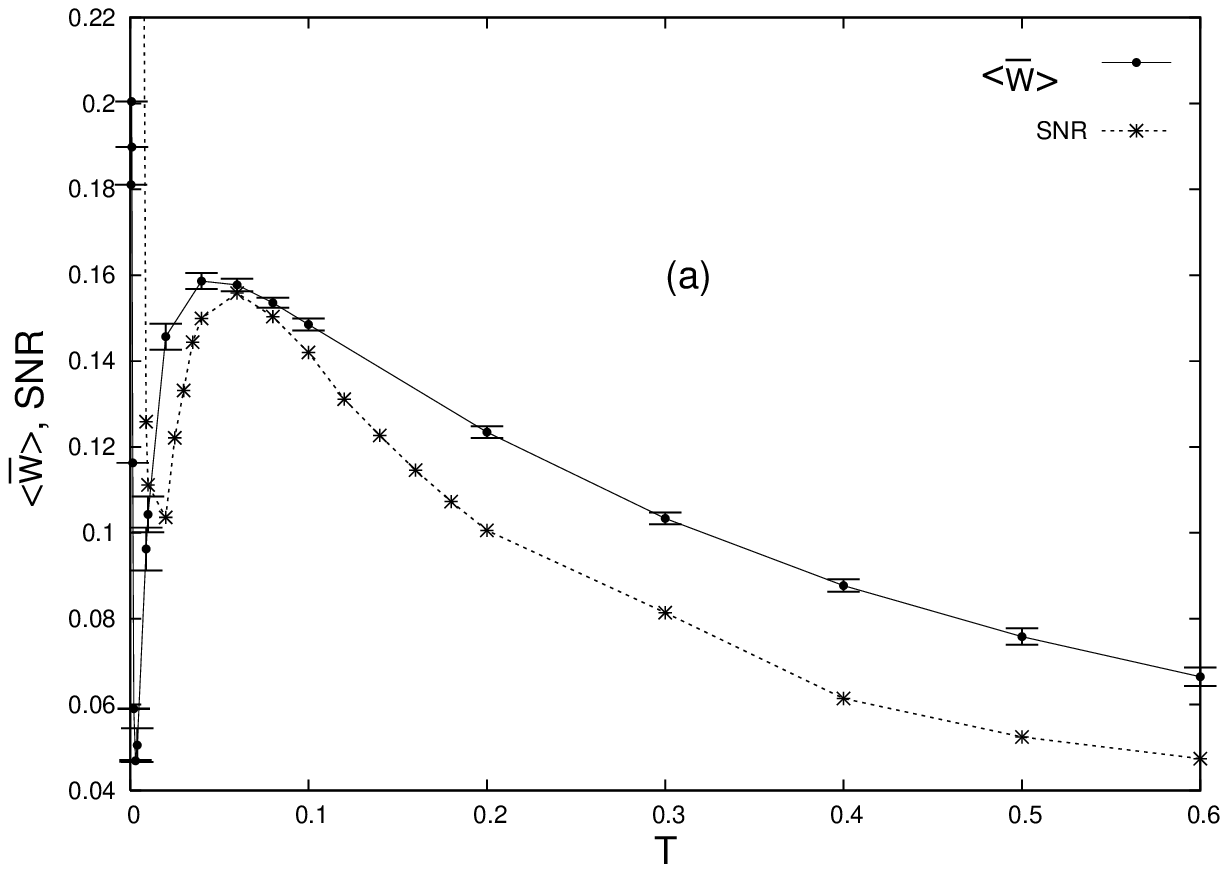}
\vspace{0.4cm}
\includegraphics[width=10cm,height=8cm,angle=0]{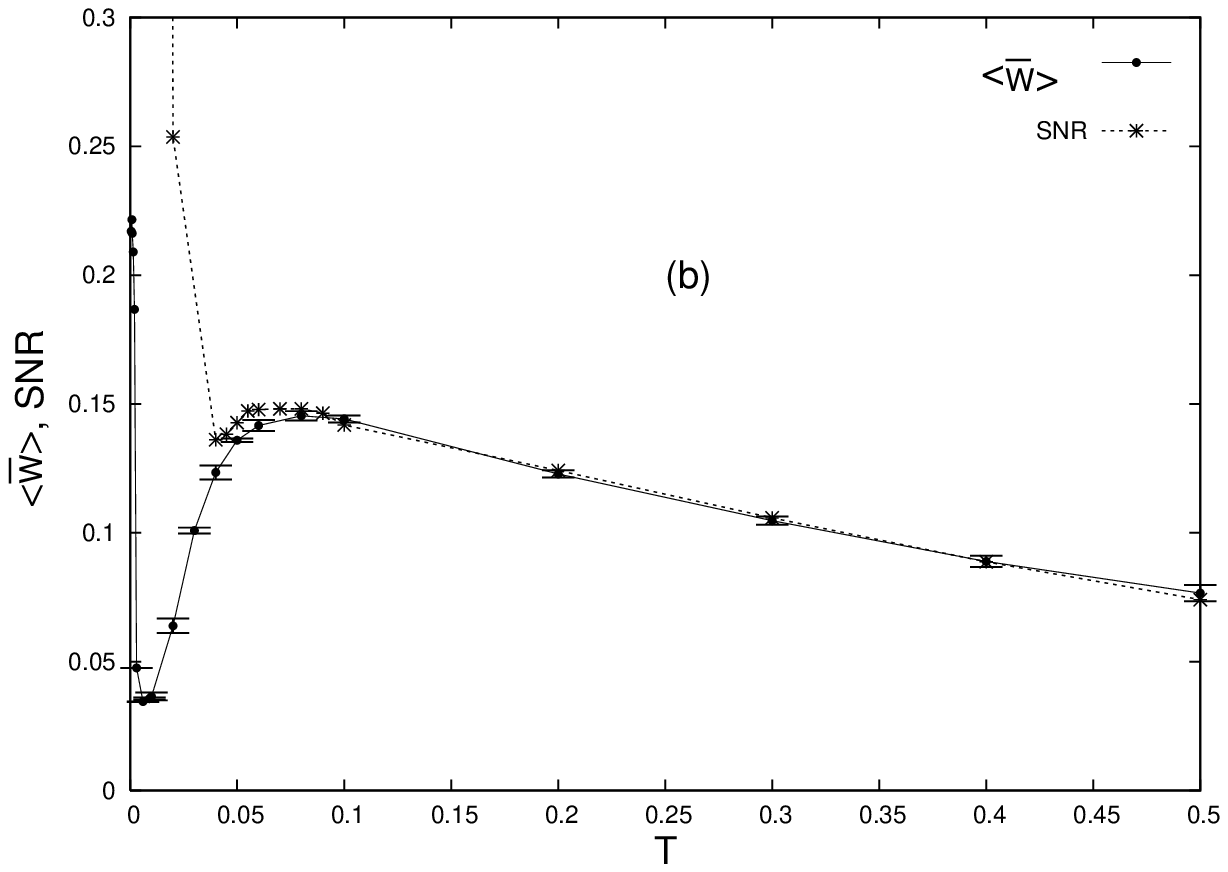}
\caption{Plot of $\langle \overline{W} \rangle$ and SNR as a function of $T$ 
for $U(x)$ (a) and $U_2(x)$ (b).}
\end{figure}

Figure 8 shows the variation of average input energy $<\overline{W}>$ or the
average hysteresis loop area $<\overline{A}>$ and the signal-to-noise ratio 
(SNR) as a function of temperature for the washboard potential $U_1(x)$. The 
same functions are plotted for the potentials
$U(x)$ and $U_2(x)$ in Fig. 9. $<\overline{W}>$ peaks at the temperatures
$T=0.2, 0.04$ and 0.08 for the potentials $U_1(x)$, $U(x)$ and $U_2(x)$, with
peak values of about 0.36, 0.16, and 0.146, respectively. These maximum values
of $<\overline{W}>$ lie between the $\overline{W}$ values of the in-phase and
out-of-phase states for the corresponding potentials. There is, however, a
remarkable difference between the nature of motion of particles at the 
temperature ($T_{SR}$) of maximum $<\overline{W}>$ for $U_1(x)$ on one hand and 
$U(x)$ and $U_2(x)$, on the other.

As the temperature $T$ goes through $T_{SR}\sim 0.2$ corresponding to maximum 
$<\overline{W}>$ , the particle feels the periodic nature of the potential 
$U_1(x)$ as the inter-well transitions are quite frequent, though not as 
frequent as the intra-well transitions between the in-phase to out-of-phase 
states. The particle motion is truly in a periodic potential implying the
presence of stochastic resonance in the periodic potential $U_1(x)$. On the
other hand, the particle has not yet begun the inter-well transitions (in 
$U(x)$) or have just started (in $U_2(x)$) as the temperature goes across the 
$<\overline{W}>$ maximum. Hence the particle does not feel the periodicity of
the potentials but the presence of the two subwells. Therefore, the resonance 
effectively occurs in a single well with two subwells. However, the bistability in a well of the potentials is required for that to happen. Moreover, the 
existence of dynamical states of trajectories is necessary for the occurrence 
of $<\overline{W}>$ maximum at that low temperature, $T_{SR}=0.04$ for $U(x)$ 
and $T_{SR}=0.08$ for $U_2(x)$.

Consider the simple Smoluchowski limit of Kramers rate\cite{Hanggi}:
\begin{equation}
k=\frac{\omega_0 \omega_b}{2\pi \gamma}\exp(\frac{-E_b}{T}),
\end{equation}
where $\omega_0^2$ and $\omega_b^2$, respectively, are the curvatures at the 
bottom of the wells (subwells) and at the top of the barrier across the 
wells (subwells) of the potential $U_1(x)$ ($U(x)$ and $U_2(x)$). This rate
calculation shows that $k^{-1}\sim 3450$ across the potential barrier between 
two consecutive wells of $U_1(x)$ ($T$=0.2, $\gamma=0.12$, $E_b=1.686$) and 
$k^{-1}\sim 47\times 10^6$ and 760 across the potential barrier between  the 
two subwells in a well of the potentials $U(x)$ ($T$=0.04, $\gamma=0.12$, 
$E_b=0.75$) and $U_2(x)$ ($T$=0.08, $\gamma=0.12$, $E_b=0.618$), respectively. 

Note that the periods $\tau=8.0, 4.8$ and 4.8 of the drive field $F(t)$ 
taken, respectively, for the potentials $U_1(x), U(x)$ and $U_2(x)$ were too 
small compared to the calculated $k^{-1}$. Therefore, the observed SR cannot 
be considered as the conventional stochastic resonance following 
Ref.\cite{Kim}. However, the resonance is brought about by the existence of 
and transition between the two dynamical (in-phase and out-of-phase) states of 
trajectories. The rates of these transitions are of the order of the period 
$\tau$ of the drive field. Viewed in this perspective of transition between 
the two dynamical states they, indeed, indicate SR. Moreover, the potential 
$U_1(x)$, at $T=T_{SR}$ a substantial number of inter-well transitions is 
observed. But in case of $U(x)$ and $U_2(x)$, inter-well transitions are 
seldom observed at $T=T_{SR}$. However, inter-subwell transitions are 
comparable in number to the transitions between the dynamical states at 
$T_{SR}$. SR nature of these transitions is further supported by the behaviour 
of input energy distributions P(W) across the temperature ($T_{SR}$) of maximum 
$<\overline{W}>$.

\subsection{Input energy distribution and SR}

\begin{figure}[htp]
  \centering
\includegraphics[width=9cm,height=6cm,angle=0]{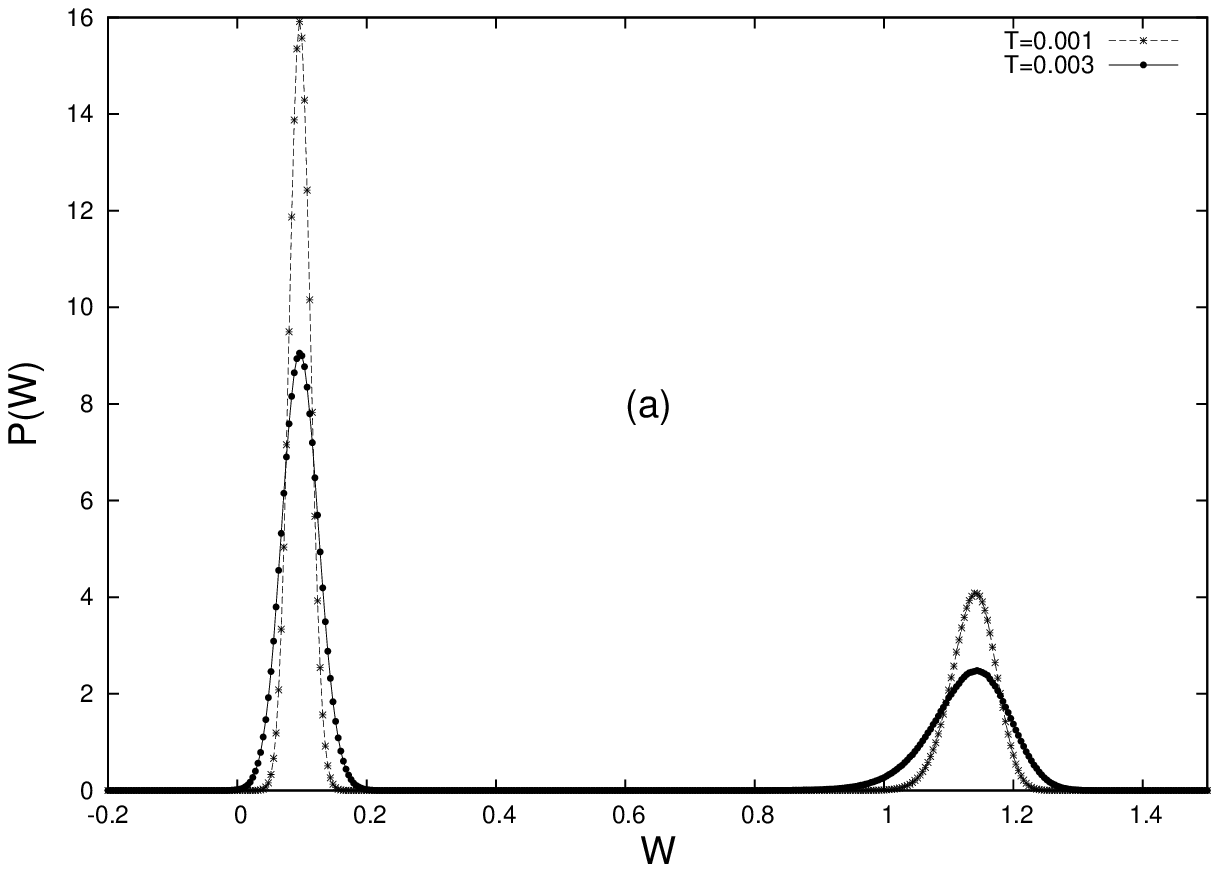}
\vspace{0.4cm}
\includegraphics[width=9cm,height=6cm,angle=0]{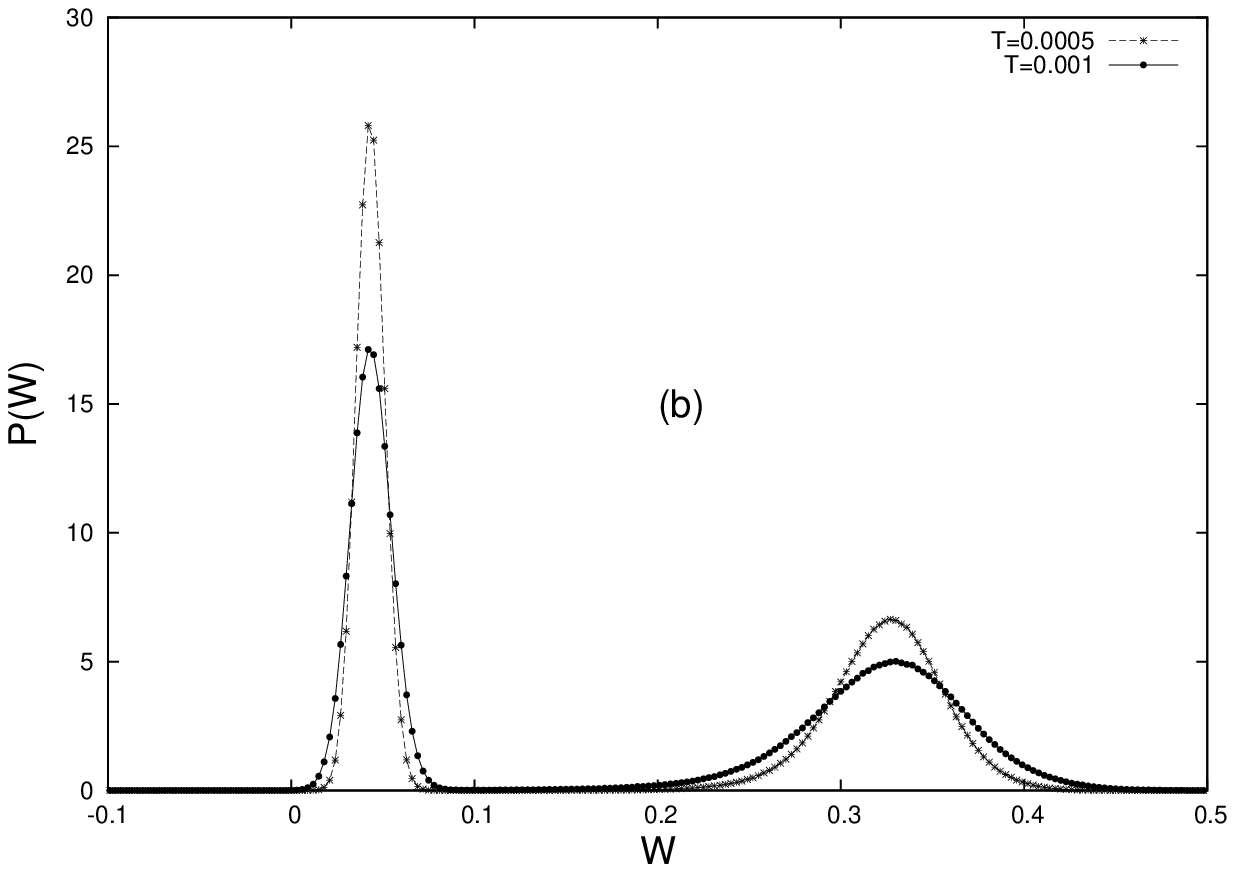}
\vspace{0.4cm}
\includegraphics[width=9cm,height=6cm,angle=0]{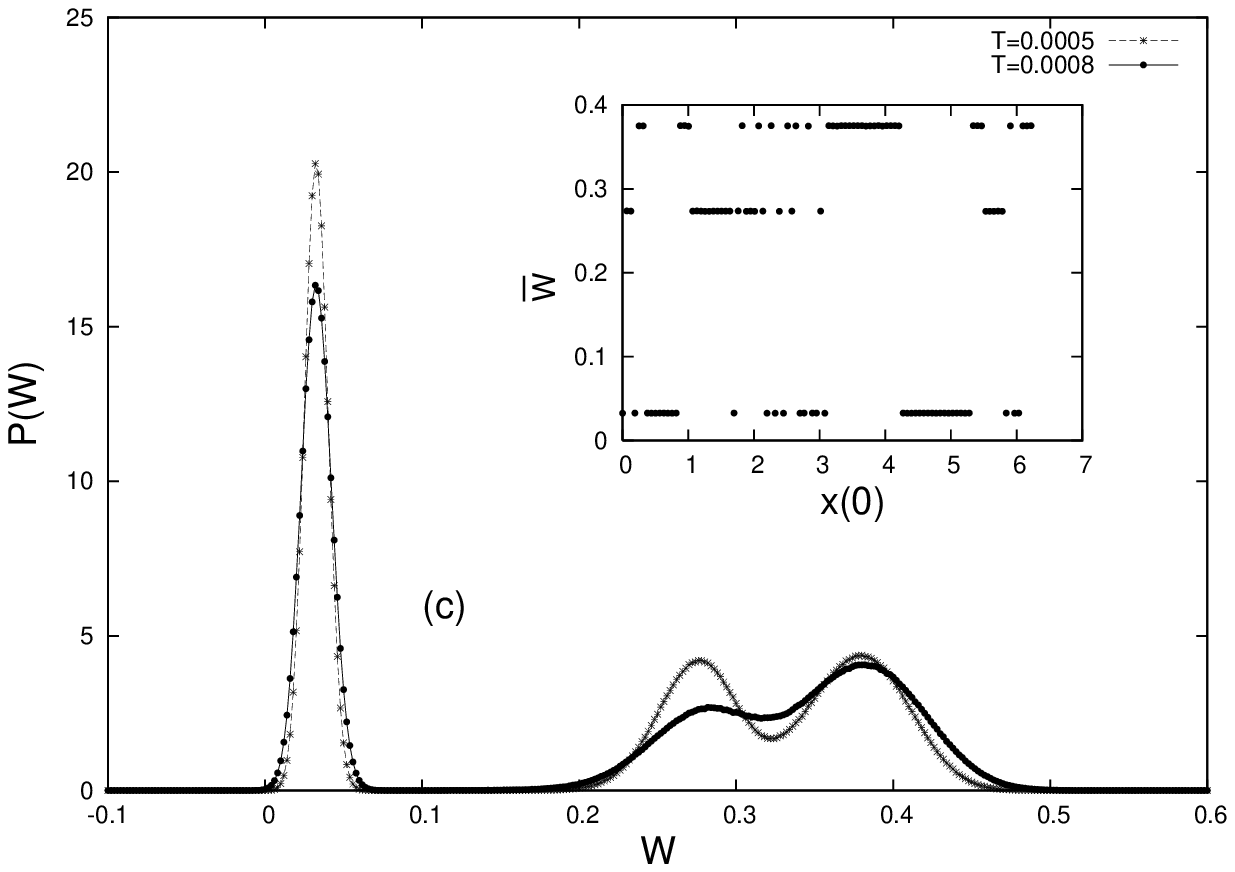}

\caption{Plot of $P(W)$ at different values of $T$ for $U_1(x)$ (a), $U(x)$ 
(b), and $U_2(x)$ (c). For small $T$, the distribution
is bimodal for $U_1(x)$ and $U(x)$. However, $P(W)$ shows three peaks for 
$U_2(x)$.}
\end{figure}

The input energy distributions $P(W)$ have earlier been used in the discussion
of SR\cite{Saikia, Sahoo, Jop, PRE}. $P(W)$, in the present case are shown in 
Fig. 10. At low temperatures, say $T=0.0005$, $P(W)$ shows the usual bimodal
distribution for the potentials $U(x)$ and $U_1(x)$. The peaks 
occur exactly at $W=\overline{W}$ values shown in Fig. 5, corresponding to the
$\overline{W}$ in the in-phase and out-of-phase states. Fig. 10 also shows 
$P(W)$ for the potential $U_2(x)$ exhibiting three peaks for the amplitude
$\Delta F=0.2$ of $F(t)$. This is because, as mentioned earlier, the two 
subwells are not equivalent because of the finite average tilt $F_0=0.1$ in
the potential $U_2(x)$. On general considerations one would expect four peaks. 
The four peak $P(W)$ appears only with $\Delta F<.19$. The three peak 
$P(W)$ instead occurs because of the instablity of the in-phase state in 
the subwell-1 for $\Delta F\geq .19$.
Here again the peaks are centered at $W=\overline{W}$ values corresponding to
the three dynamical states. The correspondence of the peaks of $P(W)$ and 
the dynamical states is thus unambiguous at low temperatures.

In Fig. 11 are shown $P(W)$ at various elevated temperatures. As the 
temperature is gradually increased two important common features could be
readily observed: (i) the variation of the strength of the $P(W)$ peaks, and
(ii) the intrusion of the in-phase peak into the negative $W$ domain and 
the appearance of a long negative $W$ tail of $P(W)$. The analysis of these two 
features provides a better understanding of SR in the three potential systems.

\begin{figure}[htp]
  \centering
\includegraphics[width=9cm,height=6cm,angle=0]{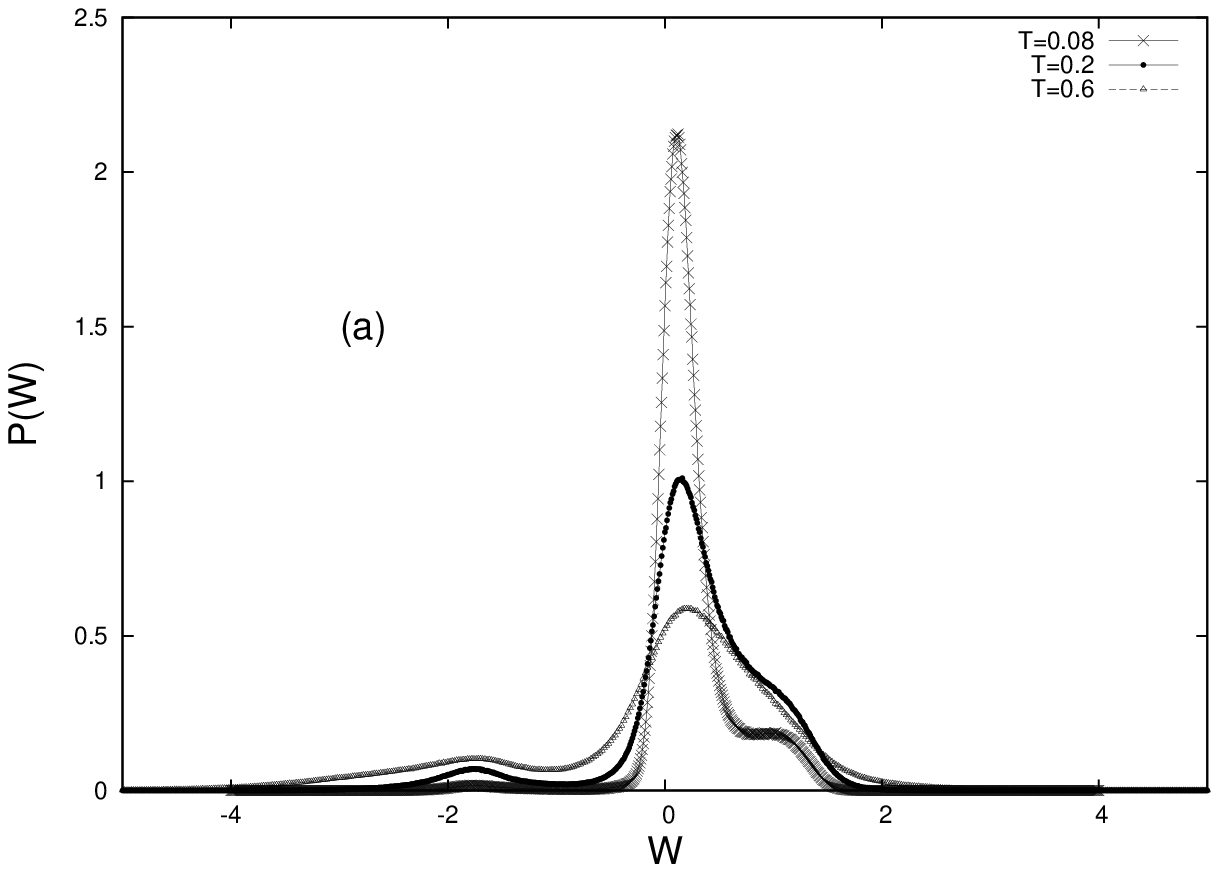}
\vspace{0.4cm}
\includegraphics[width=9cm,height=6cm,angle=0]{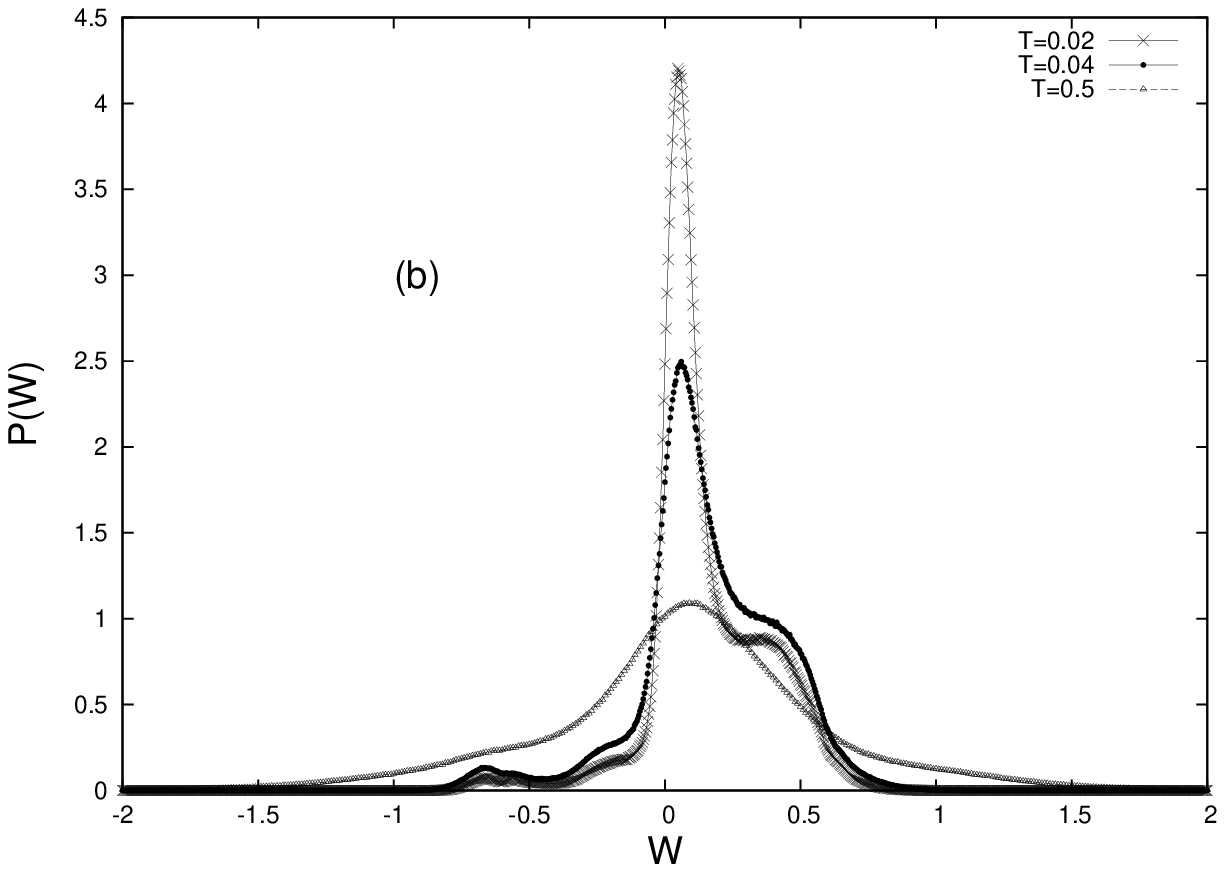}
\vspace{0.4cm}
\includegraphics[width=9cm,height=6cm,angle=0]{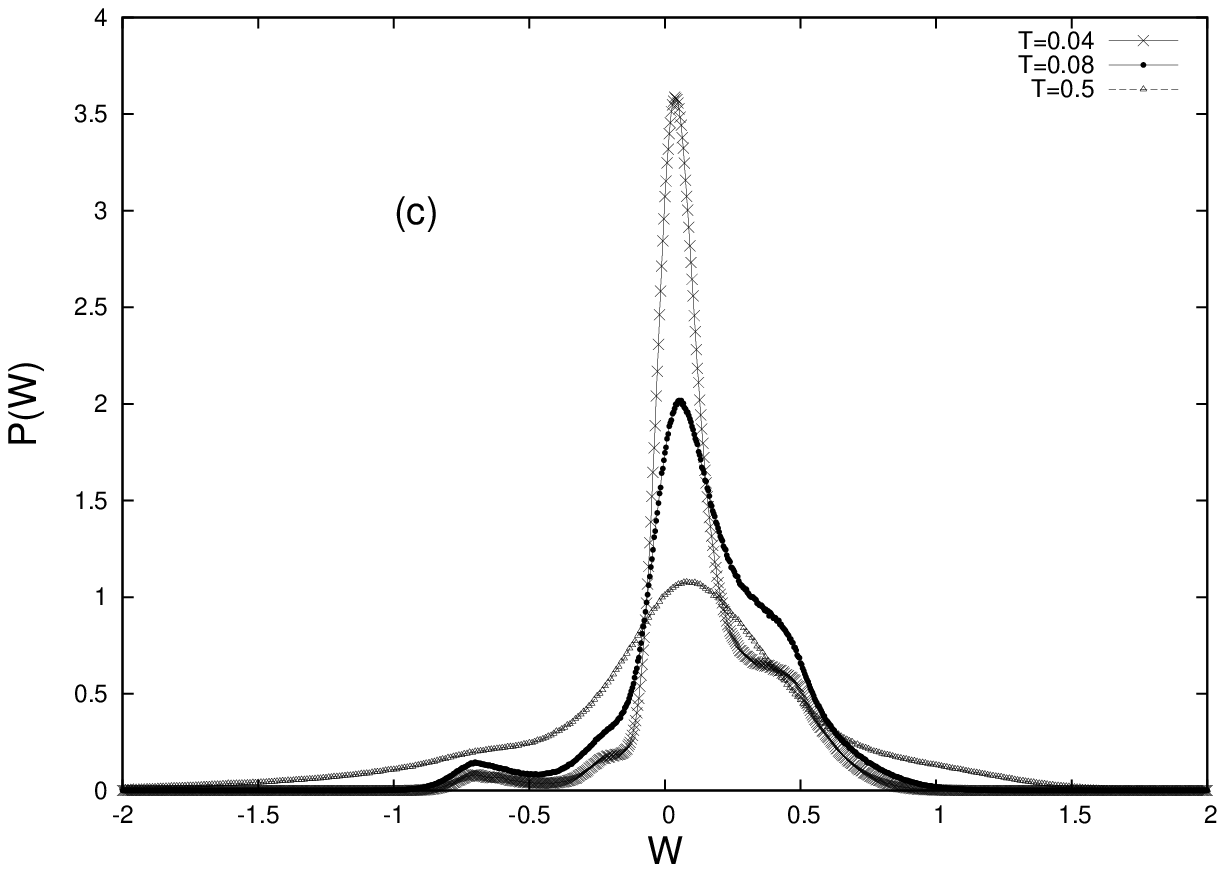}

\caption{Plot of $P(W)$ at different values of $T$ for $U_1(x)$ (a), $U(x)$ 
(b), and $U_2(x)$ (c).
At $T_{SR}$, $P(W)$ is on the verge of losing its multimodal character; 
at higher $T$ the distribution has a single peak structure.}
\end{figure}

For the potentials $U_1(x)$ and $U(x)$ the higher energy peak (corresponding 
to the out-of-phase state) first diminishes, disappears and then reappears, as 
the temperature is gradually increased. For $U_2(x)$, the peak of $P(W)$ 
corresponding to the out-of-phase state in the left subwell, i.e. op(0,1), 
first diasappears, and then the peak corresponding to op(0,2) too diasappears. 
As the temperature is increased further a broad peak (almost a plateau) appears
roughly spanning the earlier two out-of-phase peaks. As the temperature is 
increased further the newly formed peaks begin merging with the sole in-phase 
peak (for $\Delta F=0.2$) and at the temperature $T_{SR}$ the out-of-phase peak
is left only as a receding shoulder leaving no distinguishable trace of either 
a hump or a plateau. However, in the process the strength of the in-phase 
$P(W)$ peak also diminishes. This can be seen in either of the two ways: (i) by fitting the in-phase $P(W)$ peak by a Gaussian and calculating its area and 
(ii) by finding the ratio of the number of points in the stroboscopic 
(Poincar\'{e}) plots falling in the in-phase region to the total number of 
points. 

The fraction (contribution) of in-phase states in the trajectory reduces from 
1 (at $T=T_{min}\sim 0.02, 0.003$ and 0.003 for $U_1(x)$, $U(x)$ and $U_2(x)$, 
respectively, corresponding to the respective minimum of $<\overline{W}>$) 
gradually and becomes almost equal to 0.5 at the temperature $T_{SR}$ of 
maximum $<\overline{W}>$ ($T_{SR}\sim 0.2, 0.04$ and 0.08 for $U_1(x)$, $U(x)$ 
and $U_2(x)$, respectively), Fig. 12. At this temperature, it becomes hard to 
clearly distinguish the regions of in-phase and out-of-phase states of 
trajectories, just as in the liquid-gas system at the critical temperature. 
The temperature of maximum $<\overline{W}>$ is said to fall in the region of 
kinetic phase transitions\cite{Dykman2}.

The second important feature of $P(W)$ is its intrusion into the negative $W$
region. As mentioned earlier it has two components: (i) a systematic broadening
of the in-phase peak of $P(W)$ and spilling over to $W<0$ region and (ii) the 
emergence of a long $W<0$ tail. The former happens at a temperature much lower
than the temperature at which the $W<0$ tail begins to emerge. The phase lag
$\phi_1$ in case of in-phase is small ($-\phi_1\sim 0.1\pi, 0.08\pi$ and 
$0.08\pi$, respectively for the $U_1(x)$, $U(x)$ and $U_2(x)$ potentials). 
Also the in-phase $P(W)$ peak is centered close to $W\geq 0$. As the 
temperature is increased the fluctuations in the value of $\phi_1$ occur 
leading sometime to make $\phi_1>0$. In other words, the thermal effect, makes 
$x(t)$ occasionally lead the forcing $F(t)$. Thus, a few individual HLAs 
acquire a sign opposite to what is observed in the usual case when causality is 
respected. Therefore, occasionally, $W$ becomes negative and the in-phase peak
of $P(W)$ broadens into the $W<0$ region. This process continues and becomes
more evident as the temperature is increased.

Analyses of the trajectories $x(t)$ and the accompanying hysteresis loops
$x(F(t))$ reveal that the violation of causality (leading to $W<0$) occurs
during the transitions between the dynamical (in-phase and out-of-phase) 
states. In these situations the magnitude of negative $W$ (hysteresis loop area 
with opposite sign) often becomes very large. The origin of negative tail of 
$P(W)$ lies in these transitions between the dynamical phases. Since, before 
the temperature at which $<\overline{W}>$ becomes minimum at most one 
(out-of-phase to in-phase) transition occurs in the entire history of any 
trajectory spanning about $10^5$ periods of $F(t)$ the negative tail of $P(W)$ 
does not show up till the temperature $T_{min}$ of minimum $<\overline{W}>$ 
($T_{min}\sim 0.02, 0.003$ and 0.003 for $U_1(x)$, $U(x)$ and $U_2(x)$, 
respectively). 

\begin{figure}[htp]
\centering
\includegraphics[width=9cm,height=9cm,angle=0]{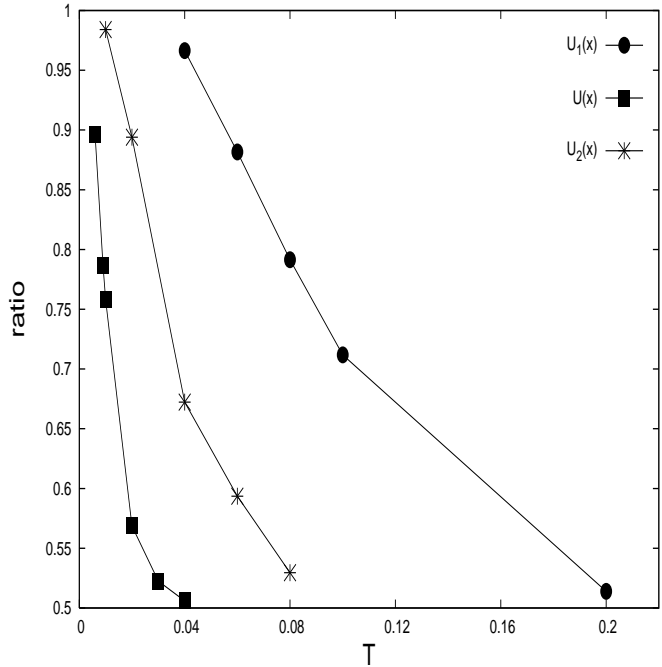}
\caption{The figure shows the plot of the fraction (ratio) of time spent by the 
particle in its in-phase state of trajectories as a function of temperature 
$T$, for the potentials $U(x)$ ($T_{SR}=0.04$), $U_2(x)$ ($T_{SR}=0.08$), and 
$U_1(x)$ ($T_{SR}=0.2$).}
\label{fig:edge}
\end{figure}

As the temperature is gradually increased further, the transitions, initially 
from the (all) in-phase to the out-of-phase states take place and  the high 
energy peak (plateau) reappears. Thereafter, transitions in both directions 
become more and more frequent ultimately causing the newly emerged high 
energy (out-of-phase) peak (plateau) to merge with the in-phase peak at the 
resonance temperature $T_{SR}$. The increase of negative tail of $P(W)$ goes 
hand in hand with increasing $T$. At temperatures $T\geq T_{SR}$ the 
transition regions dominate over the regions of in-phase and out-of-phase 
dynamical states. At these temperatures the increasing long negative tail and 
the ever broadening in-phase peak of $P(W)$ bring down $<\overline{W}>$ from 
its maximum value at ($T_{SR}$). Note that on the average causality is always 
respected and $<\overline{W}>$ never becomes negative.

\subsection{Average amplitude and phase of hysteresis loops}
\begin{figure}[htp]
  \centering
\includegraphics[width=9cm,height=6cm,angle=0]{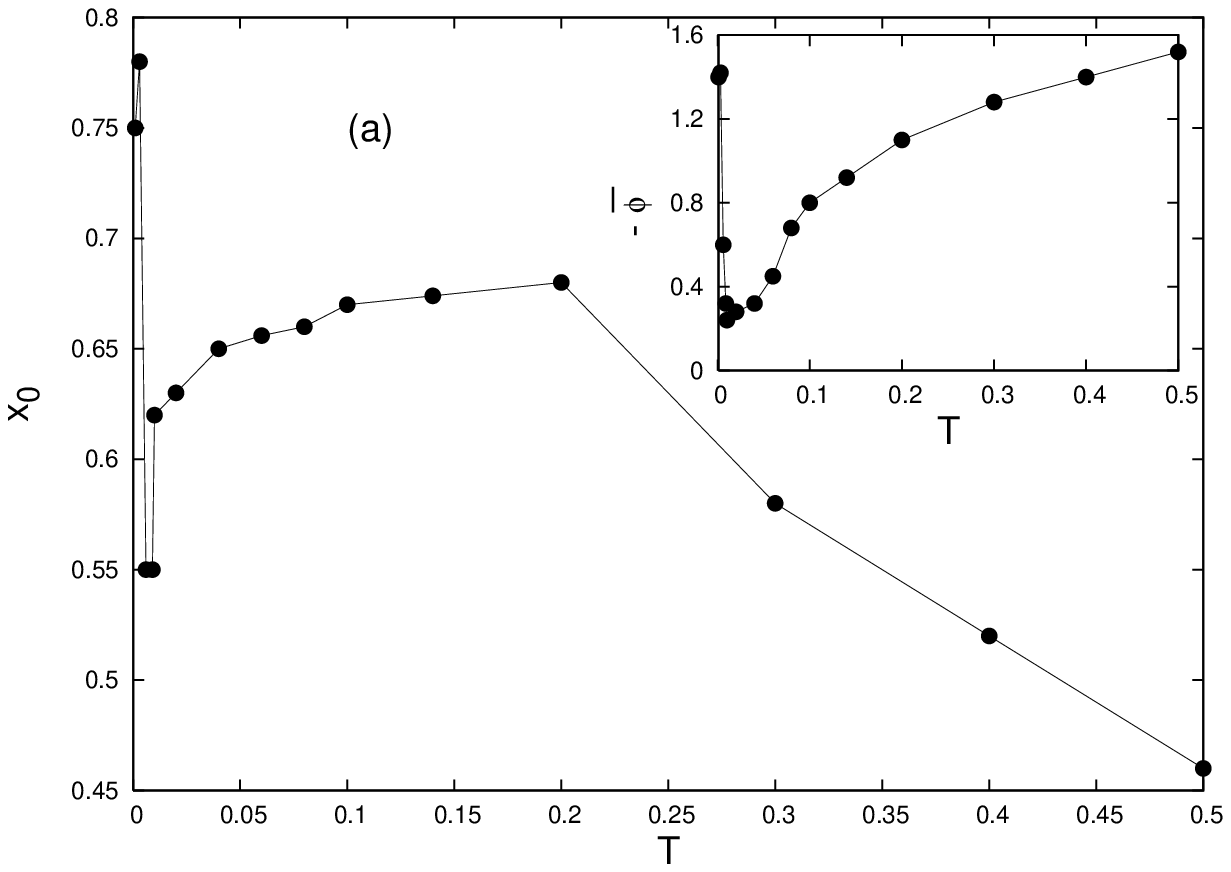}
\vspace{0.4cm}
\includegraphics[width=9cm,height=6cm,angle=0]{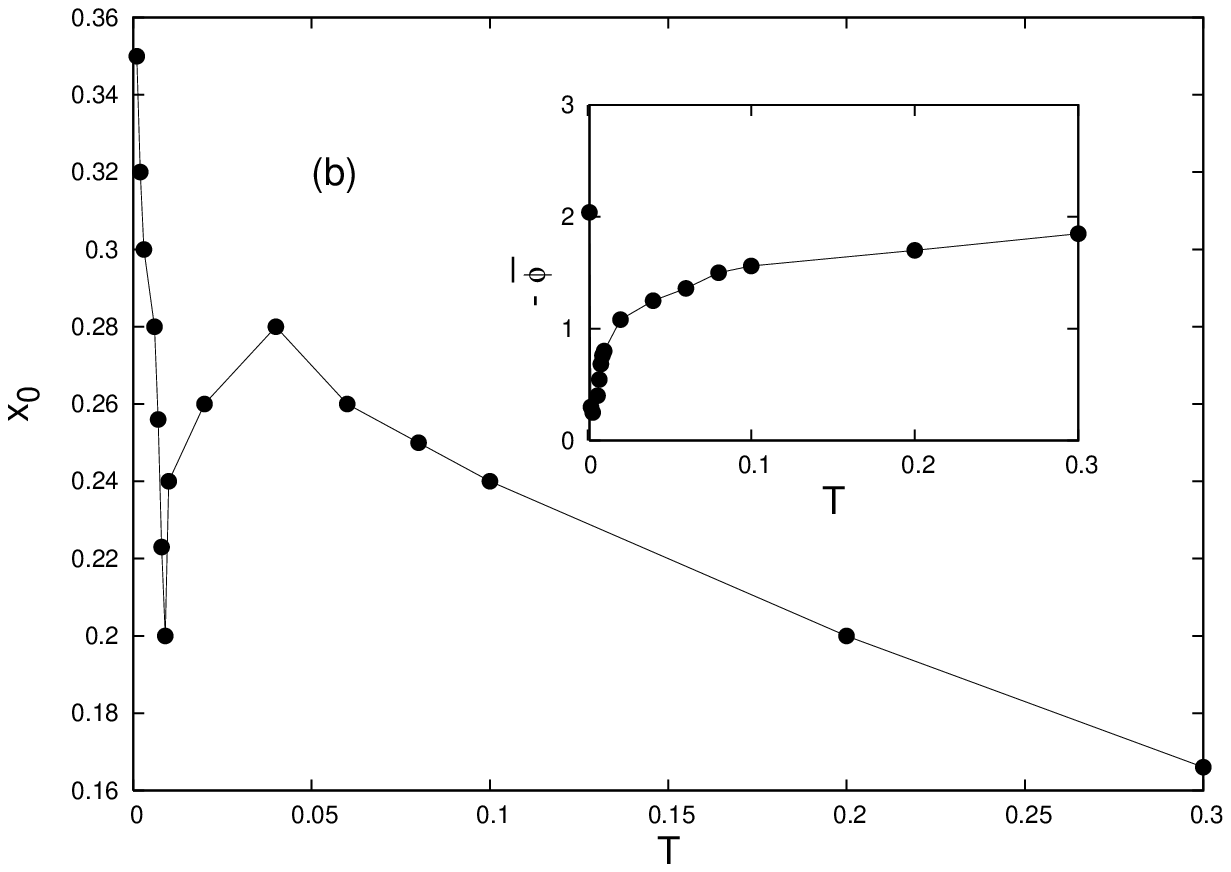}
\vspace{0.4cm}
\includegraphics[width=9cm,height=6cm,angle=0]{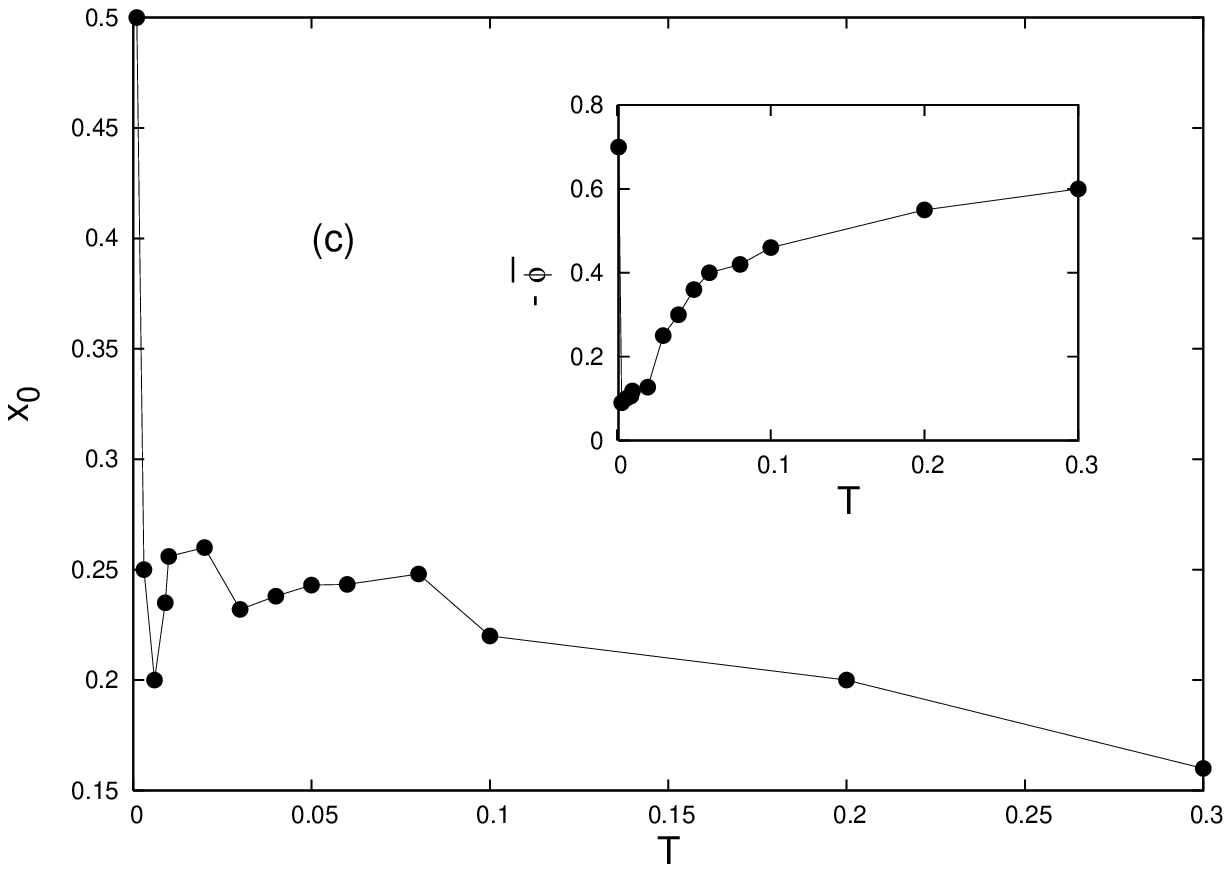}

\caption{Plot of $x_0$ and $-\overline{\phi}$ (insets)
with $T$ for $U_1(x)$ (a), for $U(x)$ (b), and for $U_2(x)$) (c).}
\end{figure}

The average hysteresis loops $<\overline{x}(F(t_i))>$ are calculated using
Eq. (3.4). Since the amplitude $\Delta F$ of the external drive $F(t)=
\Delta F\cos(\frac{2\pi}{\tau}t)$ (=0.2) is much smaller than the
periodic barrier heights ($\sim 2.0$) between two consecutive wells, the
amplitude $x_0$ of the average response $<\overline{x}(F(t_i))>$ is small and 
the relation 
$<\overline{x}(F(t_i))>=x_0\cos(\frac{2\pi}{\tau}t+\overline \phi)$, with 
$t=n\tau+t_i$, is found to follow quite well for all three potentials at low 
temperatures. However, at higher temperatures the relation serves reasonably 
well for the potential  $U_1(x)$, and only approximately for $U(x)$ and 
$U_2(x)$. 

The amplitude $x_0$ and phase $|\overline{\phi}|$ acquire their respective 
minima at the temperature $T_{min}$ of minimum $<\overline{W}>$ corresponding 
to the sole dynamical (in-phase) state, Fig. 13. At this temperature 
$\overline{\phi}\sim \phi_1$. As the temperatrure is gradually increased from
$T=T_{min}$ $x_0$ peaks at $T\sim T_{SR}$. However, $\overline{\phi}$ shows a
monotonic behaviour. At these temperatures 
$|\phi_1|<|\overline{\phi}|<|\phi_2|$ 
because the trajectories consist of a mixture of in-phase and out-of-phase 
states. The variation of $\overline{\phi}$ is similar to what is reported 
earlier\cite{LGamma} and does not show a peak\cite{Dykman1}. 

\section{Discussion and conclusion}

The existence of two dynamical states in a sinusoidally driven (under)damped
system is quite special to periodic potentials. The (quadratic) harmonic 
potentials or the (quartic) Landau potentials show only one (in-phase) state. 
The periodic nature of the potentials allows the particle to explore (in space) 
regions of high nonlinearity. The high nonlinearity of periodic potentials 
appears to be responsible for the occurrence of these two states (especially the
additional high amplitude out-of-phase state) in a well in a sinusoidal 
potential or in a subwell in the bistable periodic or washboard potential.

The investigation of dependence of the existence of two dynamical states on the
amplitude of the drive field could also be of interest in periodic potentials.
As observed earlier, the in-phase state in the left subwell of the potential 
$U_2(x)$ disappears for $\Delta F>0.19$. However, if $\Delta F$ is chosen to be
too small the particle may not have the opportunity to explore the highly 
nonlinear regions of the potential and the out-of-phase trajectories may 
disappear. Therefore, the choice of $\Delta F$ too is crucial for the study of 
SR in these potentials.  

In the conventional SR it is the bistability of the potential that plays a 
crucial role. In the present periodic (or washboard) potential case SR is 
brought about by the bistability (or multistability) of the states of 
trajectories in a well (subwell) together with the adjacent wells (subwells) of 
the potential. Moreover, if the system is initially prepared in a given well 
of a double-well potential the conventional SR occurs when the probability 
distribution of particles becomes equal in both the wells in a finite (small) 
time (typically, of the order of the period of the drive field). In other 
words, the rate of passages across the potential barrier between the two wells 
becomes equal. Exactly similar is the condition in the periodic potential 
cases considered in the present work. The ratio of time spent by the particle 
in either of the two states of trajectories to the total time reaches 0.5 as 
the temperature rises to $T_{SR}$, Fig. 12.

In order to observe SR in the `periodic' potentials discussed above it is 
necessary to choose the period $\tau$ of the drive field judiciously. SR
occurs in a narrow `window' of frequency $\omega$ ($\omega=\frac{2\pi}{\tau}$): 
approximately, [$7.8<\tau<9.8$] for the potential $U_1(x)$ and 
[$4.8\leq\tau<10.6$] for $U(x)$ and [$4.0<\tau<9.8$] for the potential 
$U_2(x)$. The lower limit of the $\tau$-window is sharp. However, the upper 
limit is not well demarcated. The $T_{SR}$ increases monotonocally with $\tau$.
Here the upper limit of the $\tau$-window is arbitrarily put so that 
$T_{SR}=0.7$. $T_{SR}=0.7$ is a large temperature and is comparable to the 
potential barrier $\sim 2$ between two adjacent wells of the potentials. 
Consequently, $<\overline{W}>(T)$ peak becomes broader with increasing $\tau$. 
However, the numerical results for the potentials $U(x)$ and $U_2(x)$ are 
expected to improve if $\tau$ is chosen larger than the value 4.8. Indeed, the 
preliminary results show that when $\tau$ is chosen somewhat larger than 4.8 
even frequent inter-well transitions could be observed around $T_{SR}$ for the 
potentials $U(x)$ and $U_2(x)$. This effectively addresses the earlier 
limitation of particle motion being confined to a single well. This provides 
further support to the thesis of the presence of SR in these `periodic' 
potentials.

MCM acknowledges partial financial support from BRNS, DAE, India under
Project No. 2009/37/17/BRNS/1959. Partial financial support from the UGC, 
India, in the form of Special Assistance Program to the Department of Physics, 
NEHU, Shillong, India is acknowledged.

\end{document}